\documentclass[superscriptaddress,showpacs,amssymb,10pt,reprint,aps,prd,longbibliography,nofootinbib,floatfix]{revtex4-1}

\usepackage{graphicx,epsfig,amssymb}
\usepackage{amsmath,amsfonts, times}
\usepackage{bm}
\usepackage{bbm}
\usepackage{epstopdf}
\usepackage[linktocpage,colorlinks]{hyperref}
\usepackage[caption=false]{subfig}
\usepackage[usenames]{color}
\usepackage{natbib}
\usepackage{soul}
\usepackage[utf8x]{inputenc}
\usepackage[usenames]{color}
\usepackage{cancel}
\usepackage{amssymb}

\usepackage{bm}
\usepackage{array}
\usepackage{booktabs}

\newlength{\CellPad}
\newcommand{\VLeft}[2]{%
\parbox[c]{#1}{%
\vspace*{\CellPad}%
\raggedright\noindent #2\par
\vspace*{\CellPad}%
}%
}

\newcommand{\VCenter}[2]{%
\parbox[c]{#1}{%
\vspace*{\CellPad}%
\centering\noindent #2\par
\vspace*{\CellPad}%
}%
}

\begin{document}
 \newcommand{\bq}{\begin{equation}}
 \newcommand{\eq}{\end{equation}}
 \newcommand{\bqn}{\begin{eqnarray}}
 \newcommand{\eqn}{\end{eqnarray}}
 \newcommand{\nb}{\nonumber}
 \newcommand{\lb}{\label}

\newcommand{\PRL}{Phys. Rev. Lett.}
\newcommand{\PL}{Phys. Lett.}
\newcommand{\PR}{Phys. Rev.}
\newcommand{\PRD}{Phys. Rev. D.}
\newcommand{\CQG}{Class. Quantum Grav.}
\newcommand{\JCAP}{J. Cosmol. Astropart. Phys.}
\newcommand{\JHEP}{J. High. Energy. Phys.}

\title{\large{Scalar-induced gravitational waves from inflation with symmetry breaking}}
\author{Chong-Bin Chen}
 \affiliation{Department of Physics, Nanchang University, Nanchang, 330031, China}
 \affiliation{Center for Relativistic Astrophysics and High Energy Physics, Nanchang University, Nanchang, 330031, China}
 \author{Jia-Xi Feng}
 \affiliation{School of Fundamental Physics and Mathematical Sciences, Hangzhou Institute for Advanced Study, UCAS, Hangzhou 310024, China}
\author{Fu-Wen Shu}
\email{shufuwen@ncu.edu.cn}
 \affiliation{Department of Physics, Nanchang University, Nanchang, 330031, China}
 \affiliation{Center for Relativistic Astrophysics and High Energy Physics, Nanchang University, Nanchang, 330031, China}
\author{Linjie Song}
 \affiliation{Department of Physics, Nanchang University, Nanchang, 330031, China}
 \affiliation{Center for Relativistic Astrophysics and High Energy Physics, Nanchang University, Nanchang, 330031, China}

\begin{abstract}
We investigate scalar-induced gravitational waves (SIGWs) in an inflationary
model with symmetry breaking, in which charged scalar fields are coupled to an
isotropic triplet of Abelian gauge fields through a kinetic
function. Such SIGWs can be enhanced to a detectable level when the mixing between the inflaton and gauge-field perturbations is sufficiently large. We find that the longitudinal mode and the charge-dependent mixing between perturbations become relevant only when
the gauge-field excitation occurs sufficiently late during inflation. In this
regime, the corresponding SIGWs are shifted to ultra-high frequencies, typically can around the GHz band. We show that the parameters characterizing the effects of the longitudinal mode
and charge-dependent mixing affect the signal in qualitatively different ways. This provides characteristic signatures for distinguishing the neutral case from the charged one through the frequency profile of the stochastic gravitational-wave background.
\end{abstract}

	\maketitle

\section{Introduction}\label{s1}

Gravitational waves (GWs) have served as a new method of probing the universe, since 2016 the ground-based interferometer LIGO has first detected GWs from binary black holes \cite{Abbott:2016nmj}. After that, GW events from astrophysical sources are
continuously observed by LIGO/Virgo, such as \cite{Abbott:2016blz,Abbott:2017gyy,TheLIGOScientific:2017qsa,Abbott:2017oio,Abbott:2017vtc,LIGOScientific:2018mvr,Abbott:2020khf,Abbott:2020uma,LIGOScientific:2020stg}. Importantly,  GWs can be generated not only from astrophysical sources but also from cosmological sources \cite{Caprini:2018mtu, Figueroa:2019paj, Tanin:2020qjw, Fomin:2022ozv} in the early universe.  
Because GWs will soon decouple and become free after generation. This implies the potential to learn information about the early universe. An earlier stage than the time when the photons decoupled with other pieces, such as cosmic phase transition, reheating, and inflation, will leave signatures throughout the universe through a relic GW background.

Typical sources of cosmological GWs include primordial fluctuations, which can generate the scalar induced gravitational waves (SIGWs)  due to the non-linearity of gravity \cite{Matarrese:1992rp,Matarrese:1993zf,Ananda:2006af,Baumann:2007zm}.
 Recently, SIGWs have attracted significant attention \cite{Saito:2008jc,Kohri:2018awv,Espinosa:2018eve,Domenech:2021ztg,Yuan:2021qgz,Inomata:2025wiv}.
  Notably, the stochastic signals observed by NANOGrav \cite{NANOGrav:2023gor, NANOGrav:2023hde}, along with data from the Chinese Pulsar Timing Array (CPTA) \cite{Xu:2023wog}, the European Pulsar Timing Array (EPTA) \cite{EPTA:2023sfo, EPTA:2023fyk}, and the Parkes Pulsar Timing Array (PPTA) \cite{Zic:2023gta}, could also potentially be explained by SIGWs \cite{Cai:2023dls, Wang:2023ost, Yi:2023mbm, Yi:2023tdk, Harigaya:2023pmw, Liu:2023ymk, Chen:2024twp, Liu:2023hpw}. 
In this paper, we investigate the SIGWs sourced by scalar fluctuations arising during inflation. Depending on their characteristic frequencies, such signals may be probed through complementary observational windows, including future laser-interferometric observatories such as LISA \cite{Danzmann:1997hm,LISA:2017pwj}, Taiji \cite{Hu:2017mde}, TianQin \cite{TianQin:2015yph}, and the Einstein Telescope \cite{ET:2019dnz}.
The MHz--GHz regime, which lies outside the sensitivity window of these laser-interferometric observatories, has also motivated proposed high-frequency GW searches based on electromagnetic and microwave resonance techniques
\cite{Li:2007ab,Li:2009zzy,Aggarwal:2020olq,Berlin:2021txa,Blas:2026ybh}.

These primordial fluctuations are usually produced during the last 50 e-folds of the inflation. Although inflation, as a phenomenological framework, has been successful in explaining the very early stage of cosmic evolution \cite{COBE:1992syq,WMAP:2003elm,Planck:2018vyg}, its details of microphysics are still in mystery. The single-field inflation is a kind of simplest model to realize inflation. In this model, inflation is driven by a single scalar field slowly rolling down a flat potential. However, the UV sensitivity of higher-order operators can lead to some issues, such as the $\eta$-problem and the super-Planckian problem that violate the flatness of potential \cite{Baumann:2014nda,Lyth:1996im,Baumann:2011ws}. It is proposed that other fields can play an important role during inflation to resolve these issues. For example, introducing an extra scalar field to undertake the large replacement in field-space \cite{Brown:2017osf,Christodoulidis:2018qdw,Renaux-Petel:2015mga,Bjorkmo:2019fls}. 

In this paper, we instead consider gauge fields. We consider a kinetic coupling $f^2(\phi)F^2$ to counteract the dilution of gauge fields in an expanding universe if $f\propto a^{-2}$. Attractors of this kind of models reveal a constant electric field during inflation \cite{Watanabe:2009ct,Kanno:2010nr,Murata:2011wv,Maleknejad:2012fw}. Herein, interesting phenomena is presented, such as statistic ansiotropy of Cosmic Microwave Background(CMB) \cite{Watanabe:2010fh,Watanabe:2010bu,Bartolo:2012sd,Himmetoglu:2009mk,Gumrukcuoglu:2010yc,Chen:2014eua}. However, they only concern the case where the ratio of the velocity of gauge fields and the velocity of the scalar field, which is denoted as parameter $h$, is small. Recently, \cite{Chen:2022ccf,Chen:2023bcz} found that if a nonzero gauge field is switched on with $h\gg 1$ during inflation, the curvature power spectrum will experience exponential enhancement near horizon-crossing. Then it is possible to use indeed GWs to detect it. 

We perform the perturbation theory of the model discussed in \cite{Emami:2010rm,Emami:2013bk,Firouzjahi:2018wlp}, which contains a complex scalar field charged under U(1) gauge fields. We assume that the gauge fields have a tiny bump with $h\gg1$ over a very short period of time such that the curvature power spectrum exhibits sufficiently large and oscillating overdensity at small scales. These scalar fluctuations acting as sources of GWs, can lead to some interesting features on the power spectra of GWs . In particular, if the bump occurs at the late time of inflation, the generated GWs will reside in an ultra-high-frequency band ($\sim\text{GHz}$). The charge coupling and the longitudinal mode become important during this late stage of inflation. These effects may induce phenomena on the power spectrum of ultra-high-frequency GWs.

The paper is organized as follows. In Section \ref{sec2}, we review inflation with symmetry breaking and present the background equations of motion in the isotropic Friedmann-Robertson-Walker (FRW) universe. Section \ref{sec3} derives the perturbations of this model and examines the characteristics of the scalar power spectrum. In Section \ref{sec4}, we focus on SIGWs during the radiation-dominated era, analyzing the enhancement of the scalar power spectrum on small scales and computing the fractional energy density of SIGWs. Our work includes three appendices, \ref{qaction}, \ref{R&Omega} and \ref{app2}, which provide related calculations in detail.

\section{Inflation with symmetry breaking}\label{sec2}
We investigate the inflation with a triplet of  charged scalar fields coupled with a triplet of U(1) gauge fields as one of the applications\footnote{We need to claim that the results in this paper are also valid for model with only one scalar and one gauge field. Although the model we discuss in this paper is more complicated, it is easier to calculate due to the isotropy of the configuration of gauge fields.}. The action of charged scalar fields is given by \cite{Firouzjahi:2018wlp}
\begin{align}
    S=&\int \mathrm{d}^4x\ \sqrt{-g}\left[\frac{M_{\text{pl}}^2}{2}R-\frac{1}{2}(\text{\textbf{D}}_{\mu}\boldsymbol{\Phi})^{\dagger}(\text{\textbf{D}}^{\mu}\boldsymbol{\Phi})-V(|\boldsymbol{\Phi}|) \right.\nonumber\\
&
  \left.  -\frac{1}{4}f^2(|\boldsymbol{\Phi}|)\text{Tr}(\text{\textbf{F}}_{\mu\nu}\text{\textbf{F}}^{\mu\nu})\right],
\end{align}
where the triplet charged scalar fields and covariant derivative are defined as
\begin{align}
    \boldsymbol{\Phi}&\equiv(\Phi_1,\Phi_2,\Phi_3)^{\top},\\
    \text{\textbf{D}}_{\mu}&\equiv\mathbbm{1}_{3\times 3}\partial_{\mu}+i\text{\textbf{e}}\boldsymbol{\mathrm{A}}_{\mu},
\end{align}
where $\text{\textbf{e}}$ is the gauge coupling constant. Here $\Phi_a$ are complex scalar fields, which are gauged by a triplet of U(1) gauge fields $A^a_{\ \mu}$ respectively. In other words, the action is invariant under the local U(1) transformation
\begin{equation}
    \boldsymbol{\Phi}\to \exp{\left(i\boldsymbol{\Theta}\right)}\boldsymbol{\Phi},\ \ \ \ \ \ \boldsymbol{\mathrm{A}}_{\mu}\to \boldsymbol{\mathrm{A}}_{\mu}-\frac{1}{\text{\textbf{e}}}\partial_{\mu}\boldsymbol{\Theta}.
\end{equation}
In this paper, we fix the symmetry by a unitary gauge in which all components of $\boldsymbol{\Phi}$ are real. Moreover, we work in an isotropic FRW universe, which is invariant under O(3) transformation. Then the configuration of background fields are chosen as $\Phi_a=\phi/\sqrt{3}$ and $A^a_{\ \mu}=(0,A(t)\delta^a_{\ i})$ for $a=1,2,3$ \cite{Bento:1992wy,Papadopoulos:2017xxx,Firouzjahi:2018wlp}. We have three gauge fields with the same amplitudes in three different spatial directions respectively, and the O(3) symmetry of spacetime can be recovered by a triple of U(1) symmetries of the gauge fields. Under this isotropic configuration, the $\boldsymbol{\mathrm{A}}_{\mu}$ acquires a dynamical mass
\begin{equation}\label{kinetic}
    (\text{\textbf{D}}_{\mu}\boldsymbol{\Phi})^{\dagger}(\text{\textbf{D}}^{\mu}\boldsymbol{\Phi})=\partial_{\mu}\phi\partial^{\mu}\phi+\frac{\boldsymbol{\mathrm{e}}^2}{3}\phi^2A^a_{\ \mu}A^{a\mu}.
\end{equation}

On the other hand, the potential and kinetic coupling are only the functions of $\phi$. Then the background equations of motion in isotropic FRW universe are \cite{Firouzjahi:2018wlp}
\begin{align}
    3M_{\text{pl}}^2H^2=&\frac{1}{2}\dot{\phi}^2+V(\phi)+\frac{3f^2\dot{A}^2}{2a^2}+\frac{\text{\textbf{e}}^2\phi^2A^2}{6a^2},\\
    M_{\text{pl}}^2\dot{H}=&-\frac{1}{2}\dot{\phi}^2-\frac{f^2\dot{A}^2}{a^2},\\
    \ddot{\phi}+3H\dot{\phi}+V_{\phi}=& \frac{3ff_{\phi}\dot{A}^2}{a^2}-\frac{\text{\textbf{e}}^2\phi A^2}{3a^2},\\
    \partial_{t}\left(af^2\dot{A}\right)=&-\frac{\text{\textbf{e}}^2\phi^2aA}{3}.
\end{align}
If the potential of the scalar field is flat enough to handle the inflation, the time-dependent mass from gauge coupling $\text{\textbf{e}}^2$ is negligible during inflation. When this gauge coupling becomes important, the inflation ends rapidly in a few e-folding numbers \cite{Emami:2010rm}. Therefore, during inflation, all gauge coupling terms can be ignored in the equations of motion and we have
\begin{equation}\label{neq}
    R_{\text{\textbf{e}}}\equiv \frac{\text{\textbf{e}}^2\phi^2A^2}{M_{\text{pl}}^2H^2a^2}\ll 1\ \ \ \ \ \ (\tau\ll \tau_e),
\end{equation}
where $\tau_e$ is the conformal end-time of the inflation.

For producing enough gauge fields during inflation, the kinetic coupling should be chosen as
\begin{align}\label{f}
    f(\phi)=\exp{\left(\frac{2c}{M_{\text{pl}}^2}\int\frac{V}{V_{\phi}}d\phi\right)},
\end{align}
where $c>1$ is a constant parameter. Then after some e-folding numbers, the energy density of the gauge fields becomes comparable to the kinetic energy of scalar fields and the system has an attractor with constant energy density of gauge fields. The attractor solution in slow-roll approximation can be presented as
\begin{align}
    \epsilon_{\phi}&\equiv\frac{\dot{\phi}^2}{2M_{\text{pl}}^2H^2}\simeq \frac{M_{\text{pl}}^2}{2c^2}\left(\frac{V_{\phi}}{V}\right)^2,\label{ephi}\\ \epsilon_A&\equiv\frac{f^2\dot{A}^2}{M_{\text{pl}}^2H^2a^2}\simeq (c-1)\epsilon_{\phi},\label{eA}\\ \epsilon&\equiv-\frac{\dot{H}}{H^2}\simeq c\epsilon_{\phi}.\label{eH}
\end{align}

We are following \cite{Emami:2010rm} to consider the symmetry breaking potential for an isotropic triplet of charged scalar fields gauged by an isotropic triplet of Abelian gauge fields
\begin{equation}
    V(\phi)=\frac{\lambda}{4}\left(|\boldsymbol{\Phi}|^2-\frac{M^2}{\lambda}\right)^2\simeq\frac{M^4}{4\lambda}-\frac{M^2}{2}\phi^2.
\end{equation}
In the approximation we have assumed that the scalar fields are far from the location of global minima of the potential $\phi\ll \mu\equiv M/\sqrt{\lambda}$ during inflation. In this paper, the inflaton starts to roll down from left to right in the potential so $\dot{\phi}>0$ and $\dot{A}<0$. The dominated slow-roll region is provided by the first term in the potential hence we have $3M_{\text{pl}}^2H^2\simeq V\simeq M^4/(4\lambda)$ during inflation. The kinetic function can be also obtained from (\ref{f}) as
\begin{align}
    f(\phi)=\left(\frac{\mu}{\phi}\right)^p,\ \ \ \ \ c=\frac{p}{p_c},
\end{align}
where the critical value of parameter $p_c$ is defined as
\begin{equation}
    p_c\equiv\frac{M^2}{2\lambda M_{\text{pl}}^2}.
\end{equation}
The slow-roll approximation $M_{\text{pl}}^2(V_{\phi}/V)^2\ll 1$ implies $p_c\gg 1$. In order to produce constant electric fields we also need $p>p_c$.

After ignoring the gauge coupling, the $\dot{A}$ can be solved from the equation of motion as \cite{Emami:2010rm}
\begin{align}\label{dAsol}
    \dot{A}=\mu\left(\frac{\xi\phi^2}{\mu^2}\right)^{p}\frac{H}{a},
\end{align}
where $\xi$ is the integration constant. Then after inserting this solution into the equation of motion of $\phi$, one can solve this equation by discarding the $\ddot{\phi}$ term in slow-roll approximation. The solution of $\phi$ in the $p\gg 1$ limit is given by \cite{Emami:2010rm}
\begin{align}
    \left(\frac{\xi\phi}{\mu}\right)^{2p}\simeq \frac{2(p-p_c)}{p^2p_c}a^4.
\end{align}
Inserting this solution into (\ref{dAsol}) and also using (\ref{ephi})-(\ref{eH}) we obtain
\begin{align}\label{eslow}
    \epsilon_{A}\simeq \frac{4(p-p_c)}{p^2},\ \ \ \ \ \ \ \ \epsilon_{\phi}\simeq \frac{4p_c}{p^2},\ \ \ \ \ \ \ \ \epsilon\simeq \frac{4}{p}.
\end{align}
One of the most important parameters in perturbation theory of this model is the ratio between the two kinetic energy of fields
\begin{align}
    h\equiv\sqrt{\frac{\epsilon_A}{2\epsilon_{\phi}}}\simeq\sqrt{\frac{p-p_c}{2p_c}}.
\end{align}
In our study, we are concerned about the large $h$ case, where the gauge fields dominate the kinetic energy of fields. In the inflation with symmetry breaking, it turns out to be $p\gg p_c$. If it is true during inflation, the field displacement is handled by the gauge fields so that the inflaton is sub-Planckian. This avoids the structure of potential on super-Planckian from higher-dimensional operators \cite{Baumann:2014nda,Lyth:1996im,Baumann:2011ws}. As we shall see, $h$ also controls the coupling of two modes during inflation. Large $h$ implies that these two modes are strongly coupled to each other even at low-energy scales.

We here derive the evolution of fields during inflation. From (\ref{eslow}) and (\ref{eH}) we have
\begin{equation}\label{phi}
    \phi^2\simeq 2M_{\text{pl}}^2p_c.
\end{equation}
On the other hand, we can solve $A$ from the equation of motion. During the attractor of constant electric fields, we have $\dot{A}\propto f(\phi)^{-2}a^{-1}$ so $\epsilon_A\propto f(\phi)^{-2}a^{-4}=$constant. Therefore, we find 
\begin{align}
    f=\Big(\frac{a}{a_e}\Big)^{-2}=\Big(\frac{\tau}{\tau_e}\Big)^2,
\end{align}
where $a_e=a(\tau_e)$ and $\tau_e$ is the end time of inflation, and for de Sitter background, $aH=-1/\tau$. Then after solving $\dot{A}$, the solution is given by
\begin{align}\label{A}
    A=\frac{2M_{\text{pl}}}{3}\sqrt{\frac{p-p_c}{p^2}}\frac{a}{f}.
\end{align}
Then we have
\begin{align}
    R_{\text{\textbf{e}}}\simeq \frac{16}{9}\frac{h^2}{(1+2h^2)^2}\left(\frac{\text{\textbf{e}}}{f}\right)^2\left(\frac{M_{\text{pl}}}{H}\right)^2.
\end{align}
$R_{\text{\textbf{e}}}\ll 1$ during inflation so that the effective gauge coupling $\text{\textbf{e}}/f$ is small during inflation to avoid the strong coupling problem.

\section{Perturbations}\label{sec3}
In this section, we treat the perturbations of this model and discuss the features of the small-scale power spectrum of curvature. After fixing the gauge symmetries and considering isotropic configurations of matter fields, We can also decompose these fields as follows \cite{Maleknejad:2011sq,Maleknejad:2011jw},
\begin{align}
\phi=&\phi(t)+\delta\phi,\ \ \ \ A^a_{\ 0}=\partial_aY+Y_a,\nonumber\\
A^a_{\ i}=&\left(A+\delta A\right)\delta_{ai}+\epsilon_{iab}\left(\partial_bU+U_b\right)+\partial_i\partial_aM\nonumber\\
&+\partial_{(i}M_{a)}+t_{ai},
\end{align}
where $\delta\phi$, $Y$, $\delta A$, $U$ and $M$ are scalar, $Y_a$, $U_a$ and $M_a$ are all transverse vector ($\partial^aY_a=\partial^aU_a=\partial^aM_a=0$), and $t_{ai}$ is a symmetric transverse traceless tensor ($\partial{^a}t_{ai}=t^a_{\ a}=0$). 

On the other hand, the perturbations of gravity are represent
\begin{align}
N=&1+\alpha,\ \ \ \ N_i=\partial_i\beta+\beta_i,\ \ \ \ g_{ij}=a^2(t)(\delta_{ij}+\gamma_{ij}),\nonumber\\
\gamma_{ij}=&-2\psi\delta_{ij}+2E_{,ij}+2W_{(i,j)}+h_{ij}.
\end{align}
Here $\alpha$, $\beta$, $\psi$ and $E$ are scalar. $\beta_i$ and $W_i$ are transverse vector ($\partial^i\beta_i=\partial^iW_i=0$) and $h_{ij}$ is a transverse traceless tensor ($\partial^ih_{ij}=h^i_{\ i}=0$). Here we are interested in the scalar perturbations $(\alpha, \beta, \psi, E, \delta\phi, Y, \delta A, U, M)$. 
Not all of them are physical. The gravitational theory has four gauge symmetries of coordinate transformations. After fixing the gauge on the scalar parts of gravity as
\begin{align}\label{gaugefix}
    \psi=E=0,
\end{align}
and eliminating the non-dynamical perturbations $(\alpha, \beta, Y)$, the remaining physical scalar is $(\delta\phi, \delta A, U, M)$, which correspond to perturbations of charged inflaton, electric fields, and magnetic fields, respectively. However, it turns out that the dominated contributions are the matter perturbations of the part of gauge fields in the gauge we chosen \cite{Watanabe:2010fh,Emami:2013bk}. Hence, we can ignore the non-dynamical gravitational perturbation $(\alpha, \beta)$. Moreover, the magnetic perturbation $U$ is decoupled with others so we can also ignore it.

\subsection{Equations of motion of the scalar perturbations}
Before writing down the equations of motion, we can define the ``longitudinal'' mode by projecting gauge fields on the direction of wavenumber 
\begin{equation}
    D\equiv-\frac{k_ak_i}{k^2}A^a_{\ i}=k^2M-\delta A.
\end{equation}
After some treatments shown in Appendix \ref{qaction}, and denoting the derivative with respect to conformal time as $'\equiv d/d\tau$, the equations of motion of scalar perturbation $(\delta\phi, \delta A, D)$ are given by
\begin{widetext}
\begin{align}
    &\delta {\phi''_c}+\left[k^2-\frac{2}{\tau^2}-\frac{8h^2}{\tau^2}\left(1+\frac{2\lambda_{\textbf{e}}}{1+\lambda_{\textbf{e}}}\right)\right]\delta\phi_c\nonumber\\
=
&-\frac{4h}{\tau}\sqrt{\frac{\lambda_{\textbf{e}}}{1+\lambda_{\textbf{e}}}}\left(\delta D'_c -\left(\frac{2}{\tau}-\frac{3}{\tau(1+\lambda_{\textbf{e}})}\right)\delta D_c\right)
    +\frac{2\Lambda_{\textbf{e}}}{3}\sqrt{\frac{1+\lambda_{\textbf{e}}}{\lambda_{\textbf{e}}}} \delta D_c
-\frac{4\sqrt{2}h}{\tau} \delta Q'_c
    +\left(\frac{8\sqrt{2}h}{\tau^2}-\frac{2\sqrt{2}\Lambda_{\textbf{e}}}{3}\right)\delta Q_c\label{eomphi}\\
&\delta{Q''_c}+\left[k^2(1+\lambda_{\textbf{e}})-\frac{2}{\tau^2}\right]\delta{Q_c}
=\frac{4\sqrt{2}h}{\tau}\delta\phi'_c+\frac{4\sqrt{2}h'}{\tau}\delta\phi_c  +\left(\frac{4\sqrt{2}h}{\tau^2}-\frac{2\sqrt{2}\Lambda_{\textbf{e}}}{3}\right)\delta\phi_c,\label{eomQ}\\
&\delta {D''_c}
    +\left[k^2\left(1+\lambda_{\textbf{e}}\right)-\frac{2}{\tau^2}+\frac{27\lambda_{\textbf{e}}}{\tau^2(1+\lambda_{\textbf{e}})^2}\right]\delta D_c\nonumber\\
    ={}& \frac{4h}{\tau}\sqrt{\frac{\lambda_{\textbf{e}}}{1+\lambda_{\textbf{e}}}}\delta\phi'_c
    +\frac{4h'}{\tau}\sqrt{\frac{\lambda_{\textbf{e}}}{1+\lambda_{\textbf{e}}}}\delta\phi_c
    -\frac{4h(5-\lambda_{\textbf{e}})}{\tau^2(1+\lambda_{\textbf{e}})}\sqrt{\frac{\lambda_{\textbf{e}}}{1+\lambda_{\textbf{e}}}}\delta\phi_c
    +\frac{2\Lambda_{\textbf{e}}}{3}\sqrt{\frac{1+\lambda_{\textbf{e}}}{\lambda_{\textbf{e}}}}\delta\phi_c, \label{eomD}
\end{align}
\end{widetext}
where the canonical variables are defined as $\delta\phi_c=a\delta\phi$, $\delta Q_c\equiv\sqrt{2}f\delta A$, $\delta D_c\equiv f\sqrt{\lambda_{\textbf{e}}/(1+\lambda_{\textbf{e}})}D$, and
\begin{equation}
   \Lambda_{\textbf{e}}\equiv \frac{a\mathbf{e}^2A\phi}{f}=\frac{4}{3\tau^2}\frac{h}{1+2h^2}\left(\frac{\mathbf{e}}{f}\right)^2\tilde{M}_{\text{pl}}^2,
\end{equation}
where $\tilde{M}_{\text{pl}}\equiv M_{\text{pl}}/H$. The parameter $\lambda_{\textbf{e}}$ is defined by
\begin{equation}\label{lambda}
    \lambda_{\textbf{e}}\equiv\frac{a^2\text{\textbf{e}}^2\phi^2}{3f^2k^2}=\frac{\text{\textbf{e}}^2\tau_e^4\tilde{\mu}^2}{3k^2\tau^6}.
\end{equation}
where $\tilde{\mu}\equiv\mu/H$.
In deriving \eqref{eomphi}-\eqref{eomD}, we have transformed the perturbations to Fourier space 
\begin{align}\label{Fs}    \delta(\tau,\boldsymbol{x})=\int\frac{\mathrm{d}^3x}{(2\pi)^{3/2}}e^{\mathrm{i}\boldsymbol{k}\cdot\boldsymbol{x}}\delta(\tau,\boldsymbol{k}),
\end{align}
where $\delta$ denotes the perturbations $\{\delta\phi_c, \delta Q_c,\delta D_c\}$ we discuss in the model, and $\delta(-\boldsymbol{k})=\delta^{\dagger}(\boldsymbol{k})$.
Note that $\lambda_{\textbf{e}}$ is not a constant during inflation. The $\dot{\lambda_{\textbf{e}}}$ can be obtained as 
\begin{align}
\dot{\lambda}_{\textbf{e}}=6H\lambda_{\textbf{e}}+\mathcal{O}(\epsilon H\lambda_{\textbf{e}})
\end{align}
by directly computing (\ref{lambda}), and the slow-roll suppressed terms have been ignored in the equations. In the early stage of inflation, we have $\lambda_{\textbf{e}}\ll 1$. In this stage, the longitudinal mode is not important compared to the electric perturbation, and therefore we can ignore $\delta D_c$ at an early time. The parameter $\lambda_{\textbf{e}}$ becomes the order unit in the late time
\begin{equation}\label{bc}
    N_c\simeq \frac{2}{3}N_e+\frac{1}{3}\log\left(\frac{k}{k_{\text{cmb}}}\frac{\sqrt{3}}{\text{\textbf{e}}\tilde{\mu}}\right).
\end{equation}
where $N_e$ is the e-folds of inflation. We see that $N_c$ increases as $k$. That is, for smaller-scale modes, the effect of charge term becomes important later. Even for the largest scale $k=k_{\text{cmb}}$, if we choose $\text{\textbf{e}}=10^{-3}$, $\tilde{\mu}=10^6$, the second term only $\sim -6$. That is, $N_c$ is usually a late time. For smaller scales (larger $k$), We obtain later $N_c$.

For small $h$, one can use the in-in formalism to compute the contribution of the power spectrum order by order. The leading contribution has been shown to be a cumulative effect on the curvature power spectrum \cite{Firouzjahi:2018wlp}. In other words, it evolves on super-horizon scales. 

For large $h$ case in our consideration, the perturbative method is unreliable, so it's difficult to track the evolution of the modes.  To calculate this power spectrum, one needs to solve the strongly mixing EoM of $\delta\phi$, $\delta Q$ and $\delta D$ numerically. However, we find that the entropy mass $m_{\delta\phi}^2/H^2\propto h^2$ is large for large $h$. When modes red-shift to low energy scales enough due to the expansion, one can integrate out the heavy mode in this system. The theory after integrating out heavy modes is an effective field theory of light modes with imaginary sound speed \cite{Chen:2023bcz}. Then, the modes are exponentially growing $\sim\exp{(-k|c_s|\eta)}$ before the horizon-crossing and the power spectrum acquires an exponential factor. 

In the next section, we will use a concrete model to solve these equations of motion numerically. We will see this exponential feature in the power spectrum of the curvature perturbation.


\subsection{Initial conditions and evolution of the perturbations}
In order to solve the equations of motion to obtain the power spectrum, we need to confirm the initial vacuum of the perturbations. 

The quadratic action in the Fourier space can be written as
\begin{align}\label{QCFs}
    S_{\text{tot}}^{(2)}=\frac{1}{2}&\int\mathrm{d}\tau\mathrm{d}^3k\Big[\Upsilon^{\dagger}{'} \Upsilon{'}-\Upsilon^{\dagger}\Omega^2\Upsilon+\Upsilon^{\dagger}{'}K\Upsilon\nonumber\\&-\Upsilon^{\dagger}K\Upsilon'\Big],
\end{align}
where $\Upsilon$ is an array of perturbations $\Upsilon=(\delta\phi_c,\delta Q_c,\delta D_c)^{\top}$, $K$ is a real and anti-symmetric matrix, 
\begin{align}
     &K_{12}=\frac{2\sqrt{2}h}{\tau},\ \ \ \ \ \ \ \ K_{13}=\sqrt{\frac{\lambda_{\textbf{e}}}{1+\lambda_{\textbf{e}}}}\frac{2h}{\tau},\nonumber\\ &K_{11}=K_{22}=K_{33}=K_{23}=0,
\end{align}
and $\Omega^2$ is a real and symmetric matrix
\begin{align}
    &\Omega_{11}^2=-\frac{2}{\tau^2}+\left[k^2-8\left(1+\frac{2\lambda_{\textbf{e}}}{1+\lambda_{\textbf{e}}}\right)\frac{h^2}{\tau^2}\right],\nonumber\\
    &\Omega_{22}^2=-\frac{2}{\tau^2}+(1+\lambda_{\textbf{e}})k^2,\nonumber\\
    &\Omega_{33}^2=-\frac{2}{\tau^2}+\left(1+\lambda_{\textbf{e}}\right)k^2+\frac{27\lambda_{\textbf{e}}}{(1+\lambda_{\textbf{e}})^2}\frac{1}{\tau^2},\nonumber\\
    &\Omega_{12}^2=-\frac{6\sqrt{2}h}{\tau^2}+\frac{2\sqrt{2}a}{3f}\text{\textbf{e}}^2\phi A,\nonumber\\
    &\Omega_{13}^2=-\frac{h}{\tau^2}\frac{6(\lambda_{\textbf{e}}-2)}{1+\lambda_{\textbf{e}}}\sqrt{\frac{\lambda_{\textbf{e}}}{1+\lambda_{\textbf{e}}}}-\frac{2a}{3f}\text{\textbf{e}}^2\phi A~ \sqrt{\frac{1+\lambda_{\textbf{e}}}{\lambda_{\textbf{e}}}},\nonumber\\
    &\Omega_{23}^2=0.
\end{align}

To determine the initial conditions of modes, we need to remove the mixing terms which are proportional to $K$, one can rotate the array of fields $\Upsilon$ as $\Phi=\mathrm{R}\Upsilon$. Here $\mathrm{R}$ is a real and orthogonal matrix that satisfies \begin{align}
    \mathrm{R}'=\mathrm{R}K,\ \ \ \ \ \ \ \ \mathrm{R}_{\text{end}}=\mathbbm{1}.
\end{align}
The second equality means that the gauge fields decay away at the end of inflation. In terms of the fields $\Phi$, the quadratic action becomes
\begin{align} \label{S-Phi}
    S_{\text{tot}}^{(2)}=\frac{1}{2}\int\mathrm{d}\tau\mathrm{d}^3 k\Big[\Phi^{\dagger}{'}\Phi'-\Phi^{\dagger}\tilde{\Omega}^2\Phi\Big]
\end{align}
with
\begin{align}
    \tilde{\Omega}^2=\mathrm{R}\left(\Omega^2+K^{\top}K\right)\mathrm{R}^{\top}.
\end{align}
where $\tilde{\Omega}^2$ is real, symmetric, and invariant under $\boldsymbol{k}\rightarrow-\boldsymbol{k}$.

At initial time, where $\lambda_{\textbf{e}}\ll 1$, we find $K_{13}/K_{12}\simeq 0$, then
 \begin{equation}
    \mathrm{R}(\tau)=\begin{pmatrix}
\cos \theta & -\sin \theta & 0 \\
 \sin \theta & \cos \theta  & 0  \\
0&0  & 1 
    \end{pmatrix}
\end{equation}
where $\theta(\tau)=\int ^\tau \left|2\sqrt{2}h/\bar{\tau}\right| \mathrm{d}\bar{\tau} $.
The details can be seen in appendix \ref{R&Omega}. From the quadratic action \eqref{S-Phi}, it is easy to get the EoM as follows
\begin{equation}
\Phi''+\tilde{\Omega}^2\Phi=0
\end{equation}
where $\tilde{\Omega}^2$ are shown in appendix \ref{R&Omega}. Then the Bunch-Davis initial condition can be chosen to satisfy $k^2\tau^2\gg \max(h^2,\tau^2\Lambda_e/\sqrt{\lambda_{\textbf{e}}})$.



\section{Scalar-induced gravitational waves }\label{sec4}
\subsection{Enhancement of scalar power spectrum on small scales}\label{sec4A}

In this paper, we mainly consider a concrete model for $h$, specifically, a Gaussian time-profile.
\begin{equation}
\label{Gau}
h(N)=h_{\text{max}}\exp \left[-\frac{(N-N_f)^2}{2\Delta^2}\right],
\end{equation}
where the profile is controlled by two parameters: $h_{max}$ and  $\Delta $. 
$N_f$ represents the feature time at which the value of $h$ reaches its maximum $h_{max}$.

We consider the switching-on of gauge fields around $N_f$ and examine the features of the power spectra at small scales for generating SIGWs. The comoving curvature perturbation is defined by $\mathcal{R}=\psi+H\delta u$, where $\delta u$ is the velocity $\delta T^0_{\ i}=(\rho+p)\partial_i\delta u$. In the spatially flat gauge $\psi=0$, the curvature perturbation can be represent as \cite{Firouzjahi:2018wlp}
\begin{align}\label{R}
    \mathcal{R}=\frac{1}{\sqrt{2\epsilon}M_{\text{pl}}}\frac{\delta\phi-\sqrt{2}h\delta Q}{\sqrt{1+2h^2}}-\frac{\text{\textbf{e}}^2\phi^2AY}{18HM_{\text{pl}}^2\epsilon}.
\end{align}
The last term is sub-dominated when $\lambda_{\textbf{e}}\ll 1$. We are interested in the Gaussian profile of $h$, and $N_f<N_e$. Hence the curvature perturbation at the end time of inflation becomes
\begin{align}\label{RQ}
    \mathcal{R}(N_e)\simeq -\frac{1}{\sqrt{2\epsilon}M_{\text{pl}}}\delta \phi(N_e).
\end{align}
We choose $N_e=60$ in our calculation\footnote{In this model, the end time of inflation usually depends on the charge $\text{\textbf{e}}$ due to the second term in (\ref{kinetic}) plays a role of mass term of the inflaton. However, in this paper we consider a short period $\Delta<\log h$ in (\ref{Gau}), which means most of time $h\ll1$ so that the period of inflation is determined by the $\dot{\phi}$.}. The effect of gauge fields on the inflaton is encoded when these gauge fields turn on during horizon-crossing. The dominated contribution to the curvature perturbation is hence from the gauge-field perturbation due to $h(N)\gg 1$ near $N_f$.

\subsubsection{\texorpdfstring{$\mathbf{e}=0$}
{Something with beta in it}}

To facilitate later comparison, we begin with the case $\mathbf{e}=0$, where the longitudinal mode $\delta D_c$  contribution is absent. We then compute the power spectra for various choices $h_{max}$ and $N_f$.

\begin{figure}[t]
\centering
\includegraphics[scale=0.36]{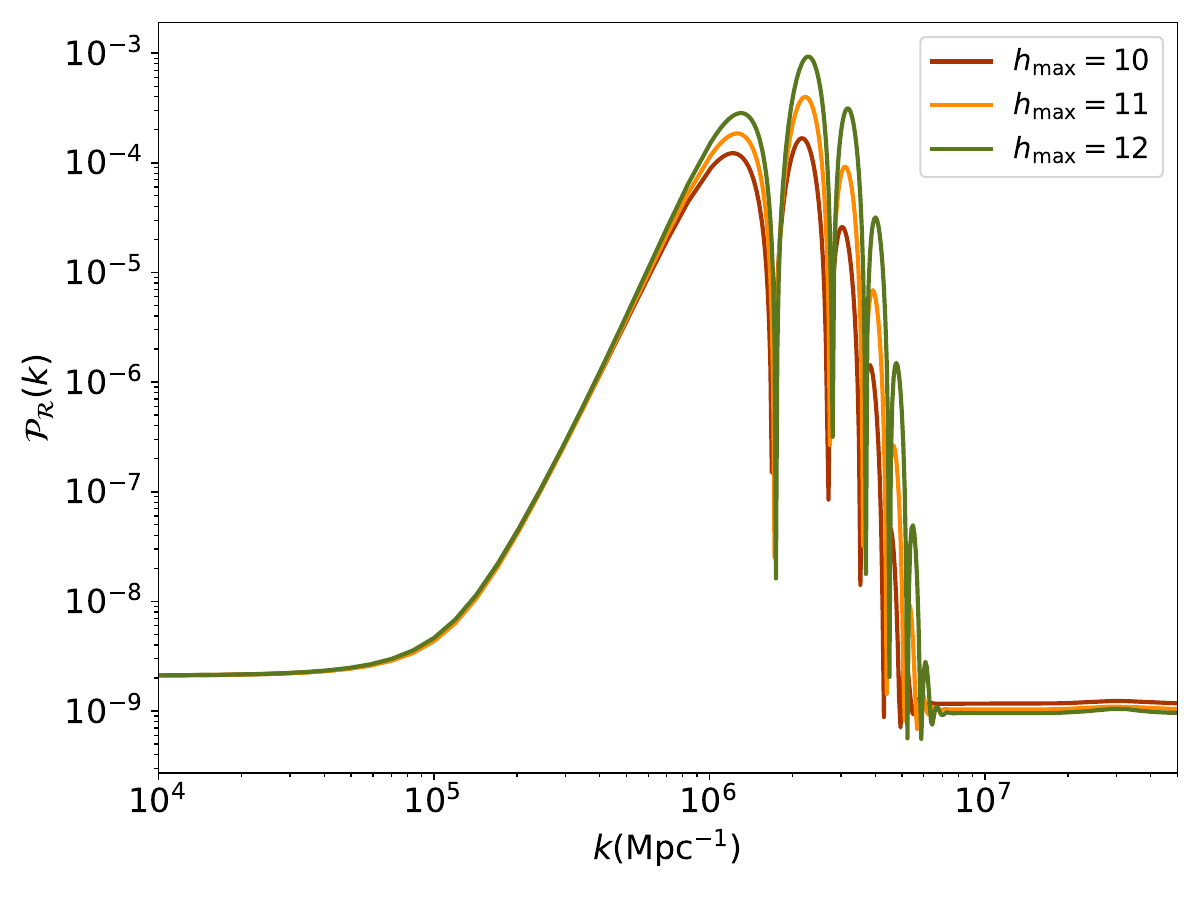}
\caption{\label{fig:h} The scalar power spectrum $\mathcal{P}_{\mathcal{R}}$ for different $h_{\text{max}}$ by fixing $\textbf{e}=0$, $\Delta=0.4$ and $N_f=15$.}
\end{figure}
\begin{figure}[t]
\centering
\includegraphics[scale=0.36]{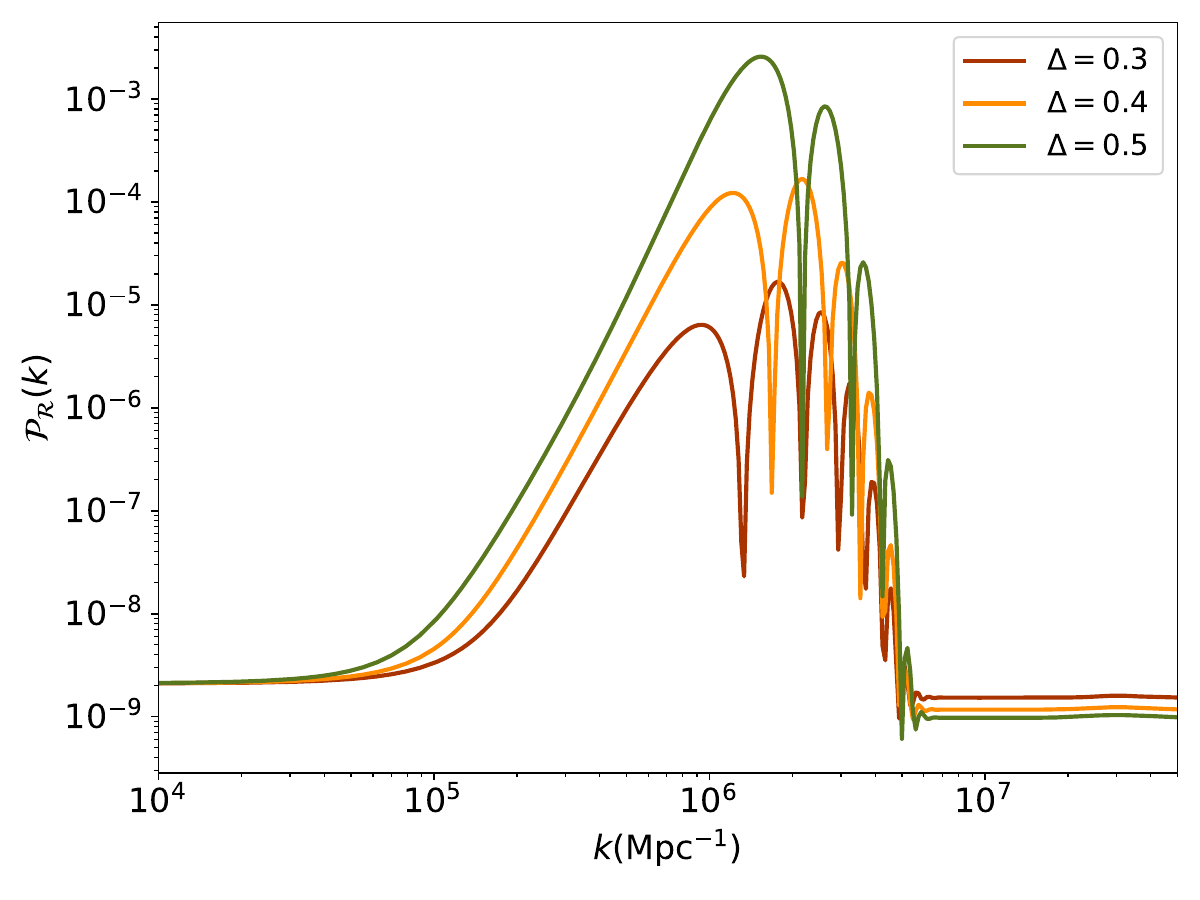}
\caption{\label{fig:delta}The scalar power spectrum  $\mathcal{P}_{\mathcal{R}}$ for different $\Delta$ by fixing $\textbf{e}=0$, $h_{\text{max}}=10$ and $N_f=15$.}
\end{figure}

Fig.~\ref{fig:h} and Fig.~\ref{fig:delta} show the scalar power spectrum $\mathcal{P}_{\zeta}$ for different $h_{\text{max}}$ and $\Delta$ respectively. We have normalized the amplitude at largest scale to fit the CMB observation $\mathcal{P}_\mathcal{R}(k_{\text{cmb}})=2.1\times10^{-9}$. A pronounced enhancement appears on the scales 
\begin{align}\label{kpeak}
    k_{\text{peak}}\sim k_fh_{max},
\end{align}
where $k_f=a(N_f)H/|c_s|$ and the amplitude of $\mathcal{P}_{\mathcal{R}}$ grows systematically with increasing $h_{max}$, especially around $k_\text{peak}$, arising from the strong coupling between $\delta\phi_c$ and $\delta Q_c$. The power spectrum is given by \cite{Chen:2023bcz}
\begin{align}\label{P0}
    \mathcal{P}_{\mathcal{R},\text{peak}}\propto e^{1.42h\Delta}.
\end{align}
For $\Delta\lesssim \log h$, the spectrum exhibits characteristic oscillatory features with multiple peaks \cite{Chen:2023bcz}. The reason is that it takes a few e-folds ($\sim \log h$) for $\delta\phi$ to grow before horizon-crossing. So, if $\Delta$ is short enough, the modes are still inside the horizon when the gauge fields have switched off, and hence they start to oscillate. This oscillation is featured by $\exp(\mathrm{i}2e^{\pm\Delta/2}k/k_f)$ \cite{Chen:2023bcz}. Therefore, a larger $\Delta$ gives a longer oscillation period, and also a higher oscillation peak, as Fig.~\ref{fig:delta} shows.

\subsubsection{\texorpdfstring{$\mathbf{e}\neq 0$}
{Something with beta in it}}

In this case, the coupling parameter $\mathbf{e}$ associated with the longitudinal mode
$D$, is taken into account. To explore the late-time behavior, we numerically solve the equations of motion (\ref{eomphi})-(\ref{eomD}) and evaluate the power spectrum for a Gaussian time profile with late turning time $N_f\gg N_c$, while fixing the model parameters ($h_{max}, \Delta$).

The influence of the charge sector is controlled by two quantities, $\Lambda_{\textbf{e}}$ and $\lambda_{\textbf{e}}$. We note that $\Lambda_{\textbf{e}}$ contains the energy scale of inflation $\tilde{M}_{\text{pl}}$, while $\lambda_{\textbf{e}}$ contains the scale of symmetry breaking $\tilde{\mu}$. So we can define
\begin{align}
    \textbf{e}_{\mu}\equiv\tilde{\mu}\textbf{e},\ \ \ \ \ \ \ \textbf{e}_M\equiv \tilde{M}_{\text{pl}}\textbf{e}
\end{align}
as parameters to discuss in the following studies.
\begin{figure}[t]
\centering
\includegraphics[scale=0.48]{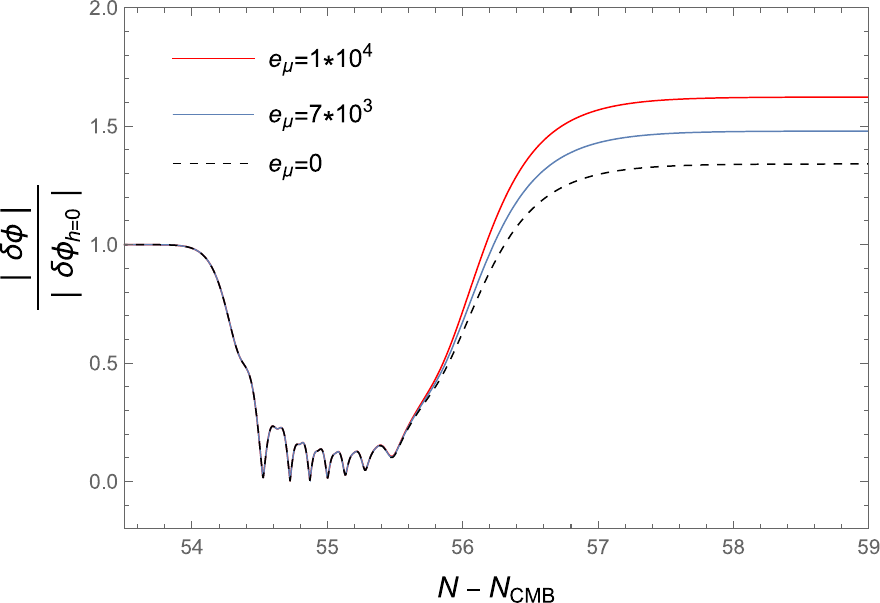}
\caption{\label{fig:evolution}The evolution of $|\delta\phi|^2$ for $k=10^{14}\text{Mpc}^{-1}(\ll k_\text{peak})$  and $\textbf{e}_M=10^2$. The parameters of profile are chosen as $h_{\text{max}}=10$, $\Delta=0.4$ and $N_f=55$.}
\end{figure}
\begin{figure}[t]
\centering
\includegraphics[scale=0.498]{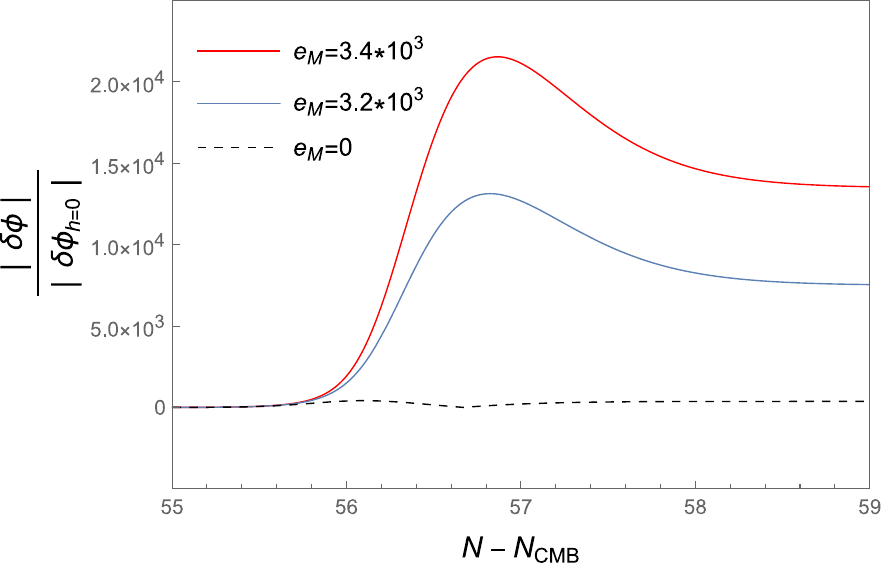}
\caption{\label{fig:evolution2}The evolution of $|\delta\phi|^2$ for $k=5\times10^{23}\text{Mpc}^{-1}(\sim k_{\text{peak}})$  and $\textbf{e}_{\mu}=10^2$. The parameters of profile are chosen as $h_{\text{max}}=10$, $\Delta=0.4$ and $N_f=55$.}
\end{figure}

Since the mode $D$ is concerned when $N_f> N_c$, all couplings, which are characterized by $\lambda_{\textbf{e}}/(1+\lambda_{\textbf{e}})$, have become constant in time $N_f$. Hence, the influence expected to be the $k^2(1+\lambda_{\textbf{e}})$ terms in the equations of motion of $\delta D_c$ and $\delta Q_c$. For $k\ll k_f$, this term becomes important when it is comparable to the Hubble friction term $2/\tau^2$. We denote $\tilde{\tau}_f$ as $k^2\lambda_{\textbf{e}}(\tilde{\tau}_f)\sim1/\tilde{\tau}_f^2$, which gives us
\begin{align}\label{tildeN}
    \tilde{N}_f\sim N_e-\frac{1}{4}\log\left(\frac{\textbf{e}_{\mu}^2}{3}\right).
\end{align}
We see that $\tilde{N}_f$ has a logarithmic dependence on $\textbf{e}_{\mu}$. That means, the mode $D$ becomes important only when the gauge fields switch on at very late time except for a very large $\textbf{e}_{\mu}$. On the other hand, for $k\gg k_f$, the influence of $\delta D$ is erased because the modes are still inside the horizon when the gauge fields are switched on. Therefore, we expect that the power spectrum is only enhanced on larger scales.

We show a late time evolution of $|\delta\phi|^2$ with different charge parameter $\textbf{e}_{\mu}$ for mode $k\ll k_{\text{peak}}$ and $\textbf{e}_M$ for mode $k\sim k_{\text{peak}}$ in Fig.~\ref{fig:evolution}. The mode evolves due to the switching on the gauge fields. We found that a larger $\text{\textbf{e}}$ would enhance the amplitude when the gauge fields are switched on. 

We also calculate the evolution with different charge parameters $\textbf{e}_M$ by fixing $k\sim k_f$ and $\textbf{e}_{\mu}$ in Fig.~\ref{fig:evolution2}. The $\Lambda_{\textbf{e}}$ term becomes important when gauge fields is switched on at time $\bar{\tau}_f$, where $\Lambda_{\textbf{e}}(\bar{\tau}_f)\sim h_{\text{max}}/\bar{\tau}_f^2$. This gives
\begin{align}\label{Nfbar}
    \bar{N}_f\sim N_e-\frac{1}{4}\log\left(\frac{\textbf{e}_M^2}{2h_{\text{max}}^2}\right).
\end{align}
For $\mu\sim M_{\text{pl}}$, $\bar{N}_f\gtrsim\tilde{N}_f$ due to the large $h_{\text{max}}$ in (\ref{Nfbar}).

\begin{figure}[t]
\centering
\includegraphics[scale=0.36]{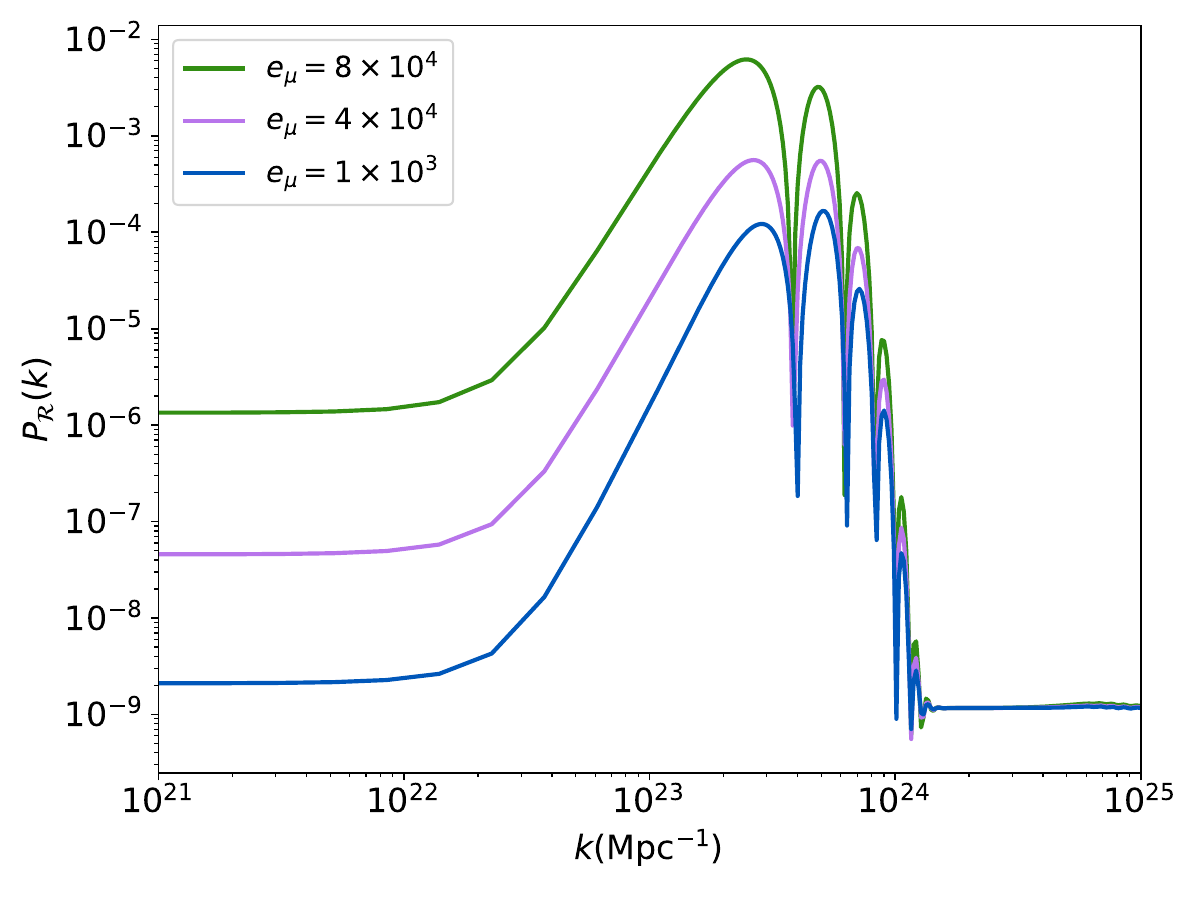}
\caption{\label{fig:PS_e}The scalar power spectrum  $\mathcal{P}_{\mathcal{R}}$ for different $\textbf{e}_{\mu}$ by fixing $\textbf{e}_M=10^2$, $h_{\text{max}}=14$, $\Delta=0.5$ and $N_f=55$. The curve of $\textbf{e}_{\mu}=10^{3}$ is renormalized to the amplitude of CMB's spectrum on large scales.}
\end{figure}

The enhancement of the power spectrum is shown in Fig.~\ref{fig:PS_e}.
If we renormalize the largest scale as the CMB's spectrum amplitude for all $\textbf{e}_{\mu}$, the peaks of the power spectrum become relatively lower since the power spectrum on large scales are raised for larger $\textbf{e}_{\mu}$. In other words, larger $\textbf{e}_{\mu}$ indicates a smaller amplitude of the peak. This is similar to reducing the value of $h_{\text{max}}$, and we will discuss it later. 

\begin{figure}[t]
\centering
\includegraphics[scale=0.36]{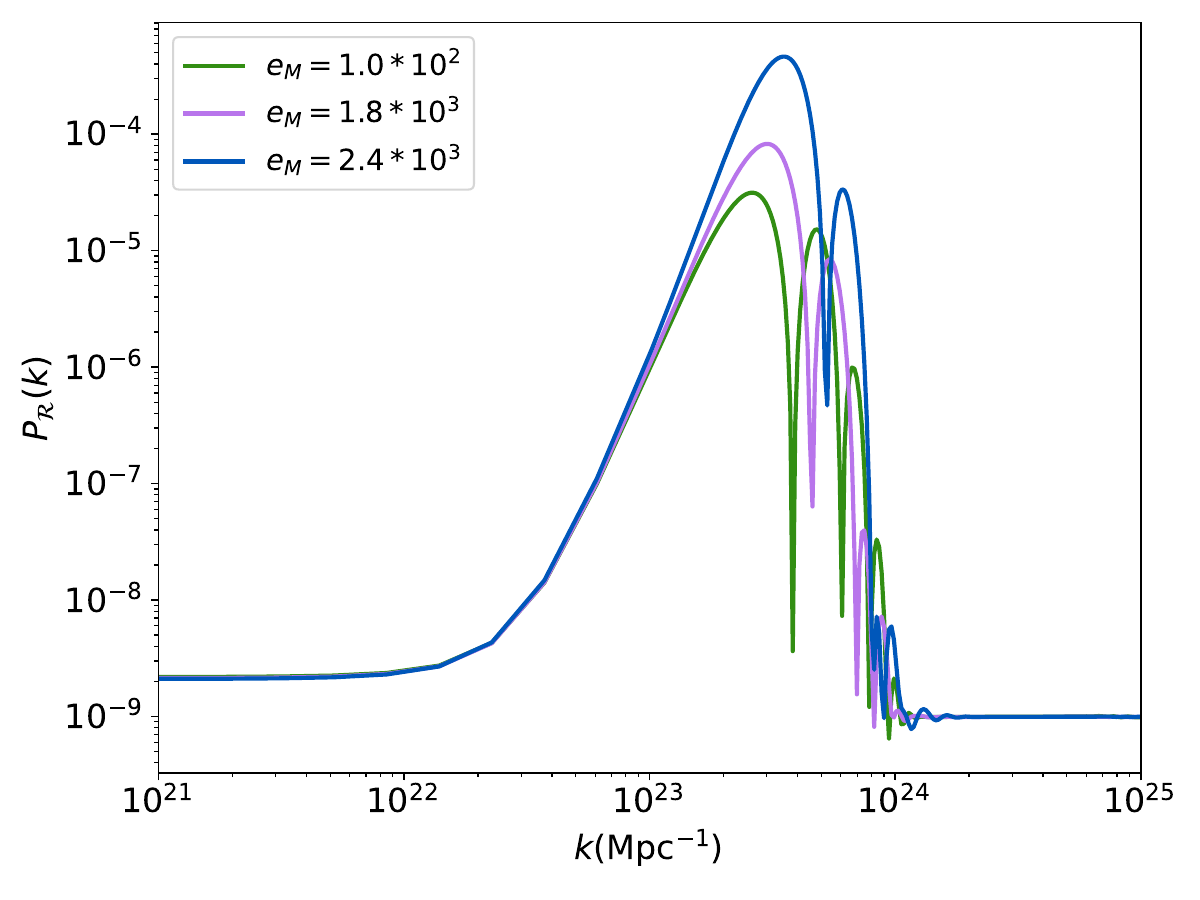}
\caption{\label{fig:PS_e2}The The scalar power spectrum  $\mathcal{P}_{\mathcal{R}}$ for different $\textbf{e}_M$ by fixing $\textbf{e}_{\mu}=10^2$, $h_{\text{max}}=8$, $\Delta=0.4$ and $N_f=55$. The curve of $\textbf{e}_{M}=2.4\times10^{3}$ is renormalized to the amplitude of CMB's spectrum on large scales.}
\end{figure}

We also find that the longitudinal mode does not affect the frequency of the oscillation features. However, raising the large-scale amplitude may change the location of the global peak. For $\textbf{e}_{\mu}=1\times10^{3}$ and $4\times10^{4}$, the global peak are given by the second peak of the oscillation. In contrastr, for $\textbf{e}_{\mu}=8\times10^{4}$, the first peak of the oscillation is raised to be higher than the second peak, so that becomes the global peak of the total spectrum.

On the other hand, changing $\textbf{e}_M$ does not change the large-scale amplitude significantly in the parameters region of interest. However, increasing $\textbf{e}_M$ increases the amplitudes of the first several local peaks, as Fig. \ref{fig:PS_e2} shows. It can also be observed that increasing $\textbf{e}_M$ can decrease the oscillation frequency. This is similar to increasing $\Delta$ in the Gaussian profile of $h(N)$.

Thus we find that $\textbf{e}_{\mu}$ and $\textbf{e}_M$ have opposite effects: the former suppresses the peak amplitude, while the latter enhances it and also lowers the oscillation frequency. We will leave a more quantitative analysis to the subsequent discussion on SIGWs.

\subsection{SIGWs sourced by scalar modes}\label{sec4B}

Based on our analysis of the scalar power spectrum in the presence of a transient gauge-field excitation,
we now compute the resulting SIGWs. These GWs are generated after inflation, when the enhanced scalar fluctuations re-enter the horizon and act as second-order sources. Throughout this section, we assume that the relevant enhanced modes re-enter the horizon during the radiation-dominated era.

The frequency of SIGWs is related to $N_f$ and $h$ as \cite{Fumagalli:2020nvq}
\begin{align}\label{fpeak}
    N_f+\log\left(h_{\text{max}}\right)\sim \log\left(\frac{f_{\text{peak}}}{\text{Hz}}\right)+37\ \ \ \ \ \ (N_f\leq\tilde{N}_f).
\end{align}

For the range $h>5$ we are interested in, if we are interested in SIGWs with peak frequencies $f_{\rm peak}<1\,{\rm Hz}$, which may be probed by LISA or PTA, the relation above implies roughly $N_f\lesssim 35$ for $N_e=60$.  
Therefore, when $h$ reaches its maximum at $N_f$, the parameter $\lambda_{\textbf{e}}\ll1$, so that the contributions from the longitudinal mode $\delta D_c$ and the charge-dependent term $\Lambda_{\mathbf e}$ can be neglected. 
By contrast, if we considered the late-time case with $N_f\gg N_c$ in order to study the possible effects of the charge coupling and the longitudinal mode. The frequency of corresponding SIGWs should be much higher than $1\,{\rm Hz}$.
\begin{figure}[t]
\centering
\includegraphics[scale=0.35]{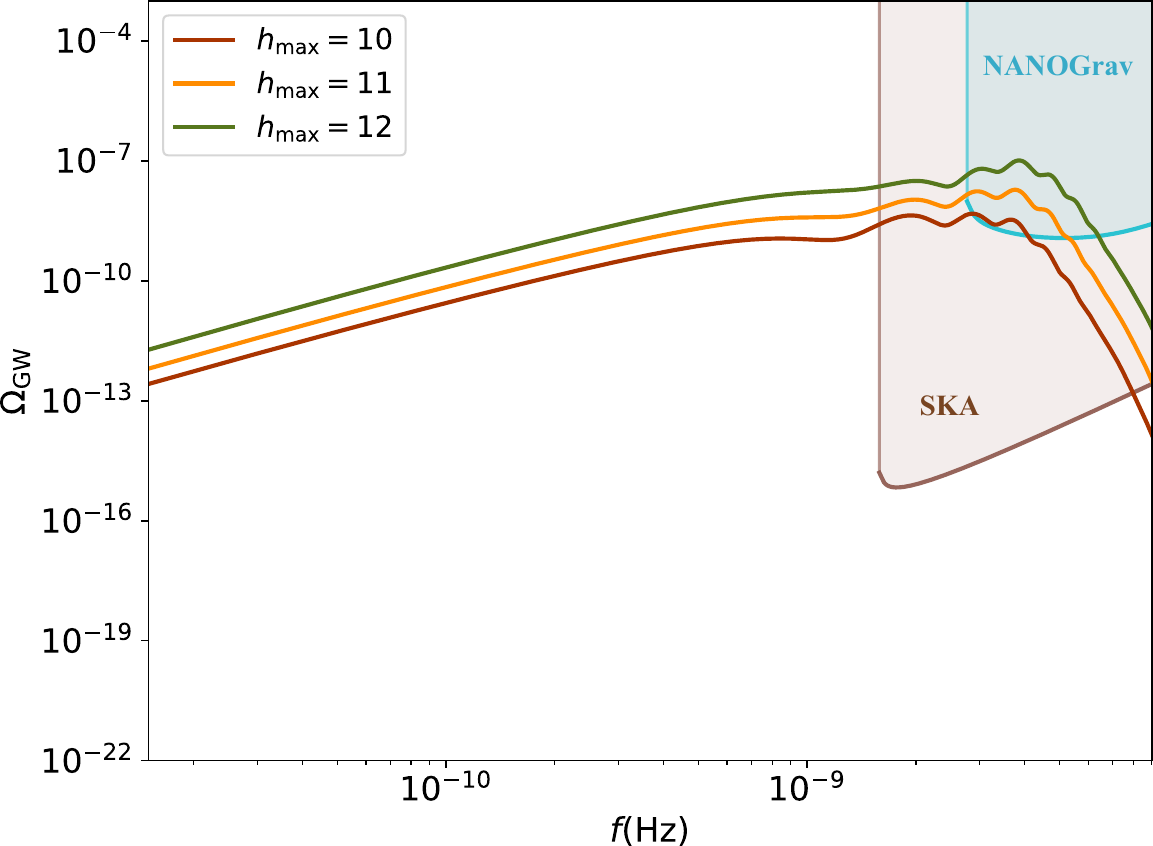}
\caption{\label{fig:GW_h}The fractional energy density of the SIGWs $\Omega_{\text{GW}}$ for different $h_{\text{max}}$ by fixing $\Delta=0.4$ and $N_f=15$.}
\end{figure}
\begin{figure}[t]
\centering
\includegraphics[scale=0.347]{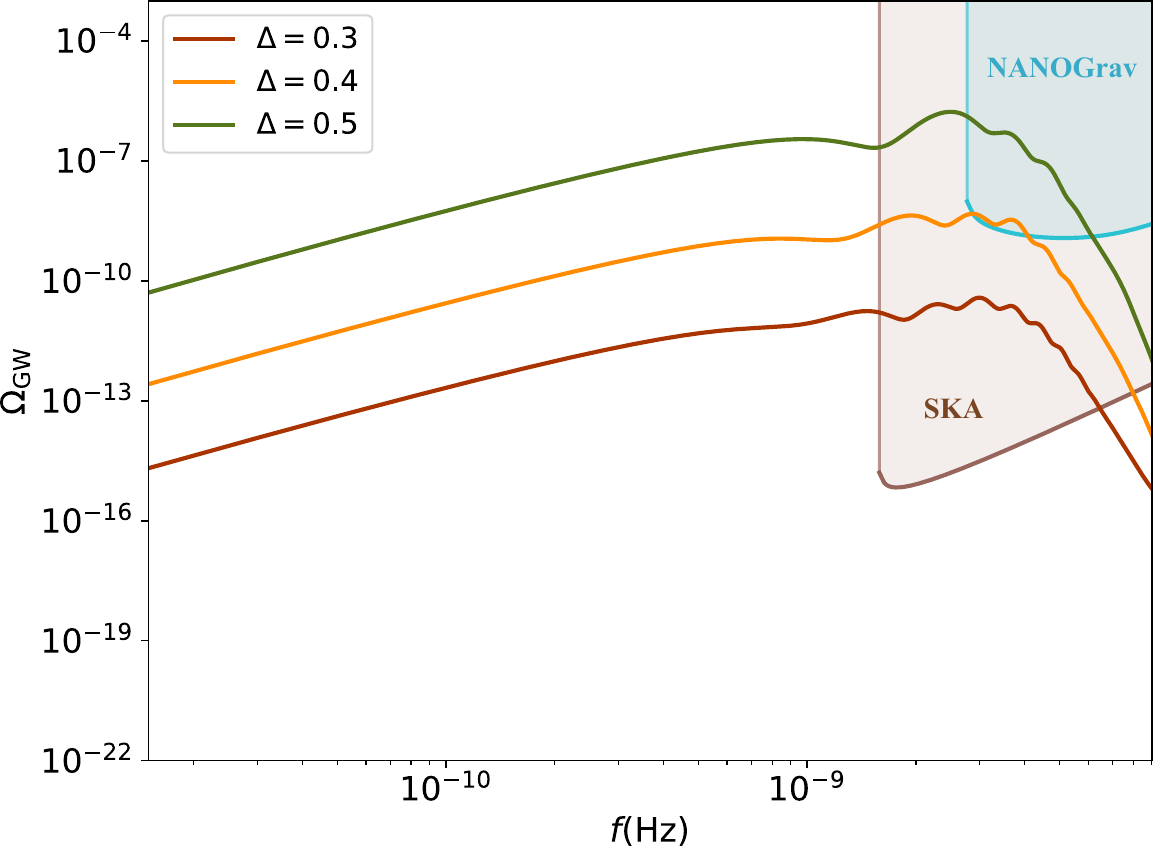}
\caption{\label{fig:GW_delta}The fractional energy density of the SIGWs $\Omega_{\text{GW}}$ for different $\Delta$ by fixing $h_{\text{max}}=10$ and $N_f=15$.}
\end{figure}

The power spectrum of SIGWs is given by  \cite{Kohri:2018awv}
\begin{align}
\label{ps-tensor}
\mathcal{P}_h(k,\tau)=\ &4\int_{0}^\infty\mathrm{d}v\int_{|1-v|}^{1+v}\mathrm{d}u
\left[\frac{4v^2-(1+v^2-u^2)^2}{4uv}\right]^2 \nonumber\\
&\times I^2(u,v,x)\mathcal{P}_{\mathcal{R}}(u k)\mathcal{P}_{\mathcal{R}}(v k),
\end{align}
where $u\equiv|\bm k-\bm{p}|/k, v\equiv p/k$, $x=k\tau$ and the kernel $I$ during radiation-dominated era is  defined as
\begin{align}
\label{I}
I(u,v,x)
=&\int_{0}^x\mathrm{d}\bar{x}~\frac{a(\bar \tau)}{a(\tau)} kG_{\bm{k}}(\tau,\bar \tau)
\frac49\Big[3 T_{\Phi}(u\bar x)T_{\Phi}(v\bar x)\nb\\
&+ \bar x \Big(\partial_{\bar \tau} T_{\Phi}(u\bar x)T_{\Phi}(v\bar x)+ \partial_{\bar \tau} T_{\Phi}(v\bar x)T_{\Phi}(u\bar x) \Big)
\nb\\
&
+\bar x^2 \partial_{\bar \tau}  T_{\Phi}(u\bar x)\partial_{\bar \tau}T_{\Phi}(v\bar x)\Big].
\end{align}
Here, the transfer function $T_{\Phi}$ and $G_{\bm{k}}$  are provided in appendix \ref{app2}.
And it's easy to obtain  that \cite{Kohri:2018awv}
\begin{align}
    \overline{I^2}= \ &\frac12\left(\frac{3(u^2+v^2-3)}{4u^3v^3 x}\right)^2\Bigg\{\Big(-4uv  \nb\\
    &+(u^2+v^2-3)\log\left|\frac{3-(u+v)^2}{3-(u-v)^2}\right|\Big)^2\nb\\
    &+\pi^2(u^2+v^2-3)^2\Theta(u+v-\sqrt{3})\Bigg\}
\end{align}
where the overline represents the time average.

The fractional energy density of the SIGWs is   \cite{Kohri:2018awv}
\begin{align}
\label{OGW}
\Omega_{\mathrm{GW}}(k,x)=&\frac{1}{24}\left(\frac{k}{\mathcal{H}}\right)^2\overline{\mathcal{P}_h(k,x)}=\frac{x^2}{24}\overline{\mathcal{P}_h(k,x)},
\end{align}
where $\mathcal{H}\equiv aH$. The GWs behave as free radiation, thus the fractional energy density of the SIGWs at the present time $\Omega_{\mathrm{GW},0}$ can be expressed as 
\begin{equation}\label{EGW}
\Omega_{\mathrm{GW},0}\left(k\right)=\Omega_{\mathrm{GW}}\left(k,\tau\rightarrow\infty\right)\Omega_{r,0},
\end{equation}
where $\Omega_{r,0}$ is the current fractional energy density of the radiation. Using eqs. \eqref{ps-tensor}-\eqref{EGW}, we plot the fractional energy density of the SIGWs generated by the Gaussian time-profile power spectrum by using a python package: SIGWfast \cite{Witkowski:2022mtg}. We show the frequency of SIGWs as $f=(1.5\times 10^{-15}k\cdot\text{Mpc})~\text{Hz}$.
\begin{figure}[t]
\centering
\includegraphics[scale=0.51]{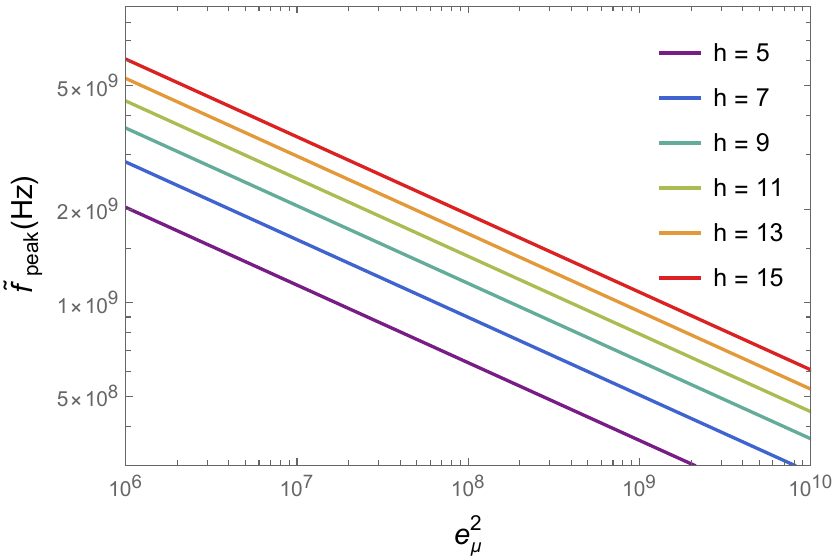}
\caption{\label{fig:tildef}Estimated peak frequency $\tilde f_{\rm peak}$ of the SIGWs affected by the
longitudinal mode as a function of $\textbf{e}_\mu^2$ for different values of
$h_{\max}$. The longitudinal-mode effect mainly appears in the GHz band.}
\end{figure}

\subsubsection{\texorpdfstring{$\mathbf{e}=0$}
{Something with beta in it}}
Without the charge $\textbf{e}$, the model is similar to the case of multi-filed inflation with hyperbolic field space. The dependence of the SIGW spectrum on $h_{\max}$ can be understood from the
corresponding scalar power spectrum. The transient gauge-field excitation enhances scalar modes around $k_{\text{peak}}$
while the scalar peak amplitude grows approximately as (\ref{P0}) Since the induced tensor spectrum is sourced by the convolution of two scalar spectra, 
$\mathcal{P}_h\propto \mathcal{P}_{\mathcal{R}}^2$. Therefore, increasing
$h_{\max}$ both shifts the GW peak to higher frequencies and enhances its amplitude rapidly. The role of $\Delta$ is twofold. A larger $\Delta$ extends the duration of the strongly
coupled phase and therefore gives the coupled $\delta\phi_c$--$\delta Q_c$ system more
time to grow. This leads to a larger scalar enhancement and, consequently, to a larger
SIGW amplitude. At the same time, $\Delta$ also changes the phase accumulated by the
modes after the gauge fields are switched off.

An exhaustive study of SIGWs produced by this transient instability of coupling with extra fields can be found in \cite{Fumagalli:2020nvq,Fumagalli:2021cel}. We summarize here several important features of this type of SIGWs:
\begin{itemize}
    \item The overall shape of the SIGW is determined by the GW signal associated with the envelope of the scalar power spectrum. 
    \item The oscillation structure in $\mathcal{P}_{\mathcal{R}}$ is inherited by the SIGWs. If the curvature power spectrum consists of periodical oscillation with frequency $\omega_{\text{lin}}$ at its peak, the modulation frequency in the gravitational-wave spectrum is related to that in the scalar spectrum by $\omega^{\text{GW}}_{\text{lin}}=\sqrt{3}\ \omega_{\text{lin}}$. The $\sqrt{3}$ factor comes from the equation of state of the radiation-dominated universe, which is equal to $1/3$. This can be observed in comparing Fig. \ref{fig:h} and Fig. \ref{fig:GW_h}.
    \item The amplitude of the modulation on the principal peak is reduced for a higher number of modulations. This is because in the integral (\ref{ps-tensor}), the modulation peaks increasingly overlap with one another so that the amplitude is averaged out. This can be observed in Fig. \ref{fig:GW_delta}.
\end{itemize}

The charged case is qualitatively different only when the feature occurs sufficiently late during inflation. when $N_f\gg N_c$, the longitudinal mode can become important, but the corresponding SIGWs are shifted to frequencies much higher than $1~{\rm Hz}$. Thus, the observable effects of the charged longitudinal sector mainly appear in the ultra-high-frequency regime.

\begin{figure}[t]
\centering
\includegraphics[scale=0.35]{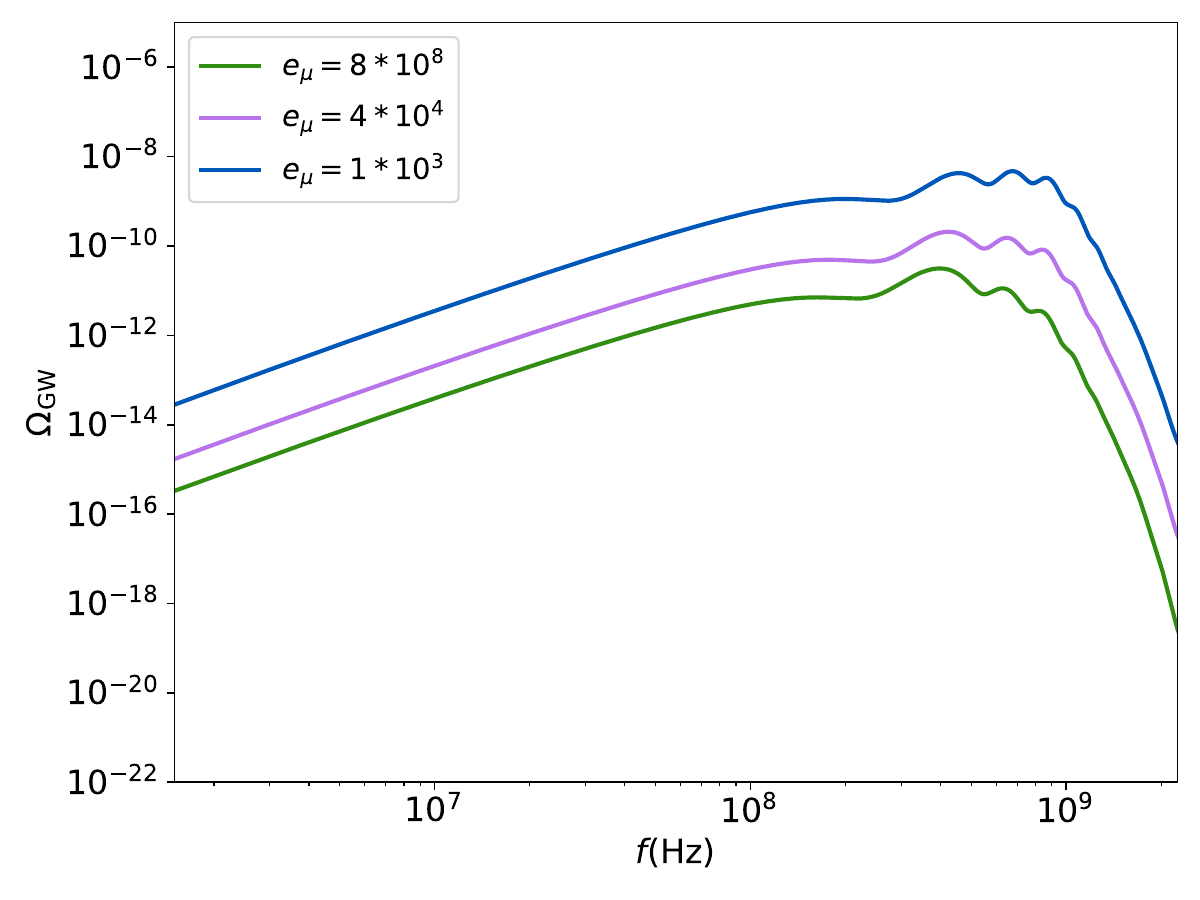}
\caption{\label{fig:GW_emu} The fractional energy density for different $\mathcal{P}_{\mathcal{R}}$ plotted in Fig. \ref{fig:PS_e}. The $\mathcal{P}_{\mathcal{R}}$ of all curves are renormalized to the amplitude of CMB's spectrum on large scales.}
\end{figure}
\begin{figure}[t]
\centering
\includegraphics[scale=0.63]{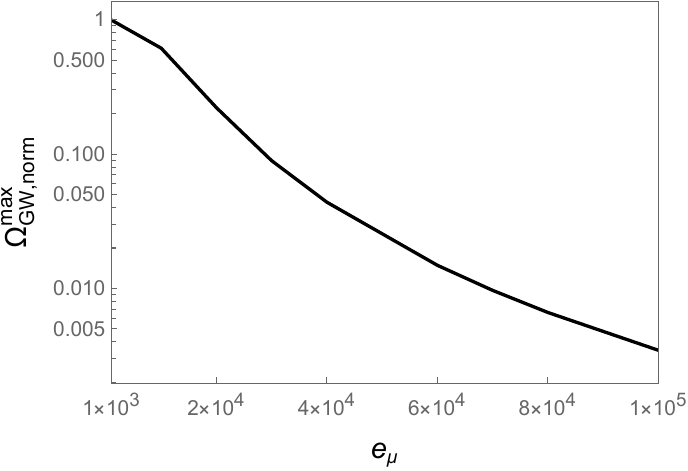}
\caption{\label{fig:GWmax}The normalized peak amplitude
$\Omega_{\text{GW,norm}}^{\text{max}}$
as a function of $\textbf{e}_\mu$. Increasing $\textbf{e}_\mu$ suppresses the relative principal peak of the SIGW spectrum. Here $h_{\text{max}}=10$ and $\Delta=0.4$.}
\end{figure}
\begin{figure}[t]
\centering
\includegraphics[scale=0.35]{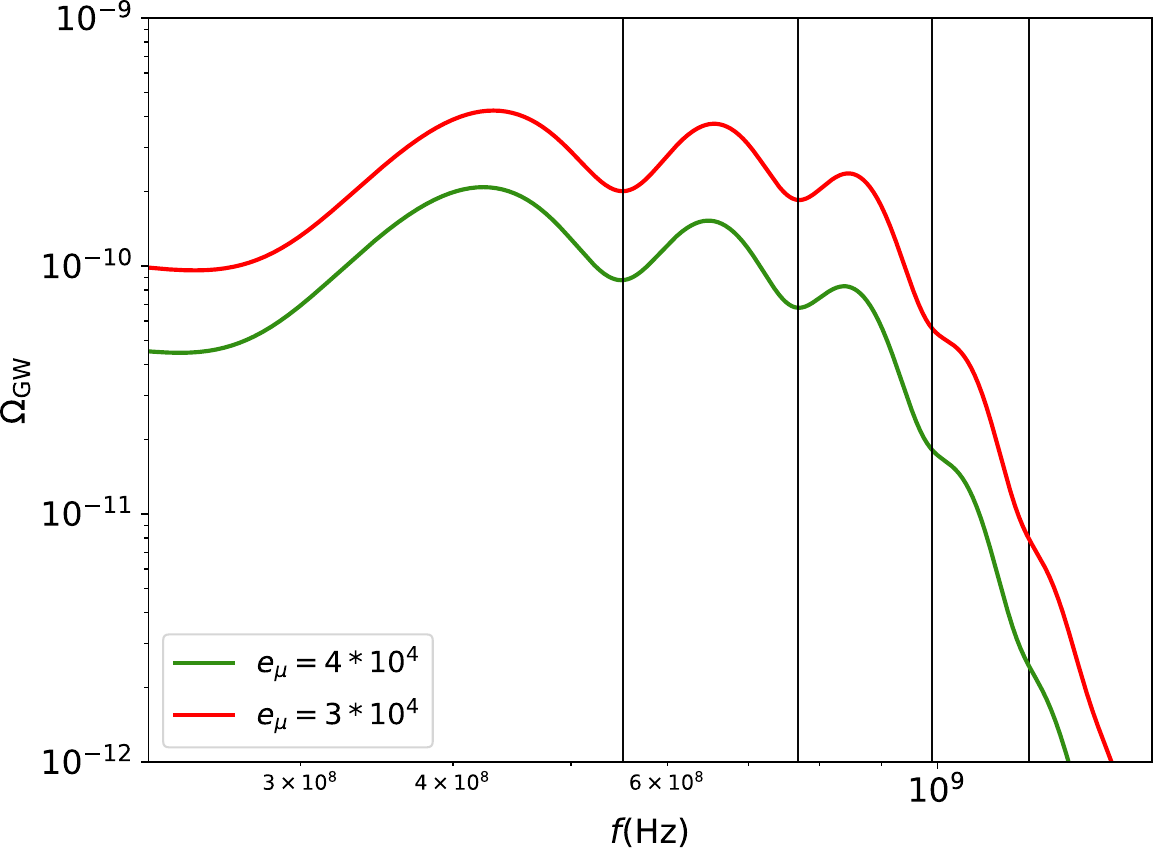}
\caption{\label{fig:GW_emu2}SIGW spectrum around the principal peak for two different
values of $e_\mu$. The vertical lines  manifest the modulation pattern. 
Changing $e_\mu$ mainly changes the relative heights of the local peaks,
while leaving the modulation pattern almost unchanged. Here $h_{\text{max}}=10$ and $\Delta=0.4$.}
\end{figure}
\subsubsection{\texorpdfstring{$\mathbf{e}\neq0$}
{Something with beta in it}}

We roughly give the ranges of the parameters $\textbf{e}_{\mu}$ and $\textbf{e}_M$ that are of interest to us. The first is the Hubble parameter. We choose $10^{-9}M_{\text{pl}}<H_{\text{inf}}<10^{-5}M_{\text{pl}}$. The upper bound of the Hubble parameter during inflation is constrained by the tensor-to-scalar ratio \cite{BICEP:2021xfz}. In general, $H_{\text{inf}}$ does not have a universal lower bound; it depends on the specific model. As we will see, the energy density of SIGWs is very sensitive to the parameters. Therefore, to avoid the parameter range spanning too broadly, we choose the lower bound coming from GUT baryogenesis \cite{Kolb:1998he}.

The parameters of in the model are $\textbf{e}$ and $\mu$. The charge $\textbf{e}$ is also model-dependent. In order to preserve the weak coupling and perturbativity of the EFT, we typically adopt $10^{-3}<\textbf{e}<1$. The symmetry-breaking scale $\mu$ can be constrained by CMB constraints for Abelian Higgs cosmic strings $\mu\lesssim 10^{-4}M_{\text{pl}}$ \cite{Lizarraga:2016onn}. Taking all of the above into account, the parameter ranges we will consider are roughly as
\begin{align}
    \textbf{e}_{\mu}<10^5,\ \ \ \ \ \ \ 10^2<\textbf{e}_M<10^9.
\end{align}
On the other hand, $N_f$ only shifts the time of switching on gauge fields, hence we fix $N_f=55$ in the rest of the discussion.

One can estimate the frequency of the corresponding SIGWs. From (\ref{tildeN}) and (\ref{fpeak}), the frequency of SIGWs produced at time $\tilde{N}_f$ is estimated as
\begin{align}
    \log\left(\frac{\tilde{f}_{\text{peak}}}{\text{Hz}}\right)\sim 23-\frac{1}{4}\log\left(\frac{\textbf{e}_{\mu}^2}{3h_{\text{max}}^4}\right).
\end{align}
We show $\tilde{f}_{\text{peak}}(\textbf{e}_{\mu}^2)$ for various $h_{\text{max}}$ in Fig. \ref{fig:tildef}. The impact of the  longitudinal mode on SIGWs is found to be primarily located in the GHz frequency band.

We first consider the effect of the longitudinal mode $\textbf{e}_{\mu}$ on the energy density of SIGWs. We have shown in the last section that increasing $\textbf{e}_{\mu}$ raises the amplitude of large-scale modes. This relatively reduced the peak of the power spectrum. Hence, it's expect that $\textbf{e}_{\mu}$ reduces the amplitude of the $\Omega_{\text{GW}}$, as Fig. \ref{fig:GW_emu} shows. The relation between $\Omega_{\text{GW,norm}}^{\text{max}}\equiv\Omega^{\text{max}}_{\text{GW}}/\Omega_{\text{GW,e=0}}^{\text{max}}$ and $\textbf{e}_{\mu}$ is calculated numerically and is shown in Fig. \ref{fig:GWmax}. We find that when $\textbf{e}_{\mu}=10^3$ the effect of the longitudinal mode can be neglected. The quantity decreases sharply with $\textbf{e}_{\mu}$, and in the large-$\textbf{e}_{\mu}$ regime it is described by a steep power law. This indicates that the observable is highly sensitive to the parameter $\textbf{e}_{\mu}$.
For example, a $10\%$ increase in $\textbf{e}_{\mu}$ suppresses $\Omega_{\text{GW}}^{\text{max}}$ by approximately $23\%$
in the large-$\textbf{e}_{\mu}$ regime.

Since the longitudinal mode does not affect the frequency of the oscillation feature of the power spectrum, the modulations of SIGWs are almost unaffected (see Fig. \ref{fig:GW_emu2}). However, under unchanged modulation, increasing the amplitude of large-scale modes shifts the global peak position. As $\textbf{e}_{\mu}$ varies, the SIGW peak position exhibits a sudden jump. These shifts can be observed in the Fig. \ref{fig:GW_emu}. This differs from simply varying $h_\text{max}$, since (\ref{kpeak}) indicates a linear dependence of the peak position on $h_{\text{max}}$.
 .

\begin{figure}[t]
\centering
\includegraphics[scale=0.35]{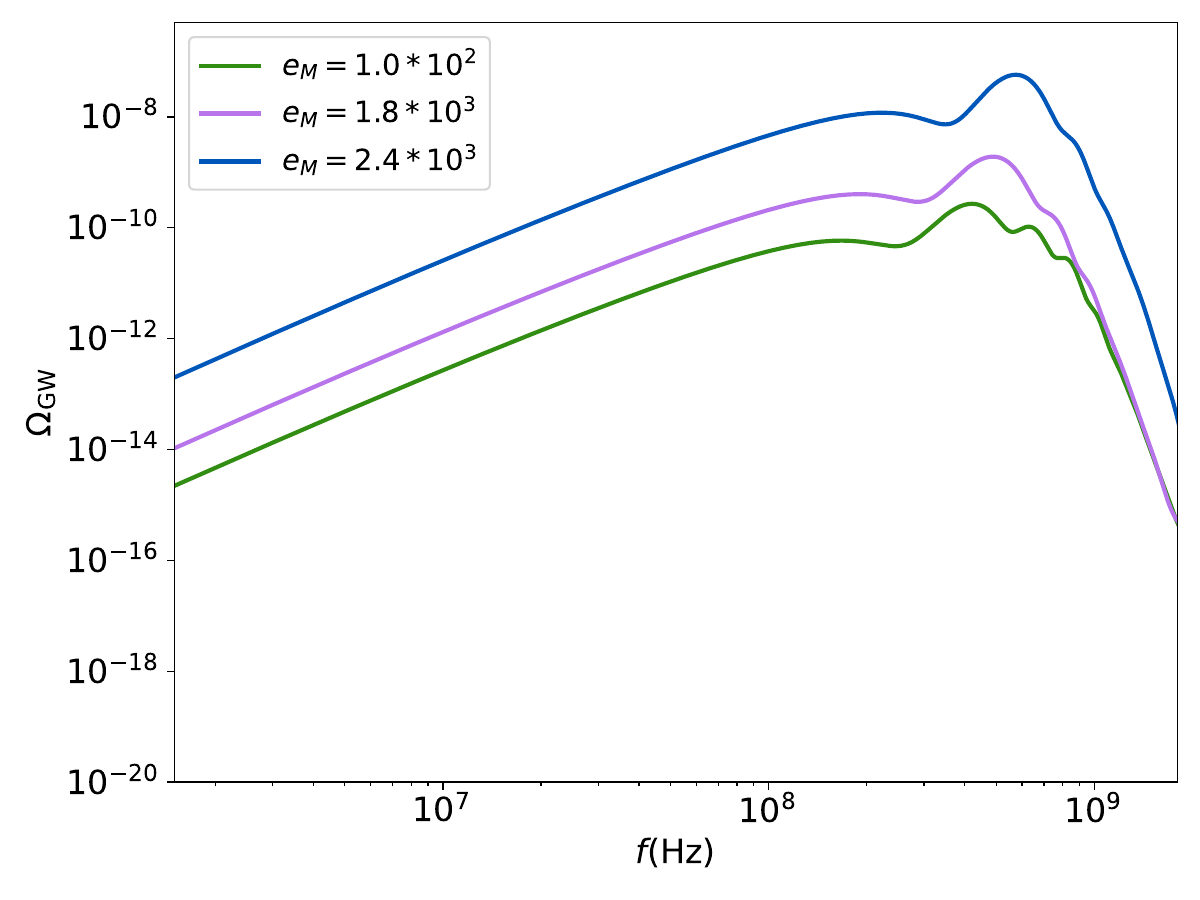}
\caption{\label{fig:GW_eM}  The fractional energy density for different $\mathcal{P}_{\mathcal{R}}$ plotted in Fig. \ref{fig:PS_e2}. The $\mathcal{P}_{\mathcal{R}}$ of all curves are renormalized to the amplitude of CMB's spectrum on large scales.}
\end{figure}
\begin{figure}[t]
\centering
\includegraphics[scale=0.6]{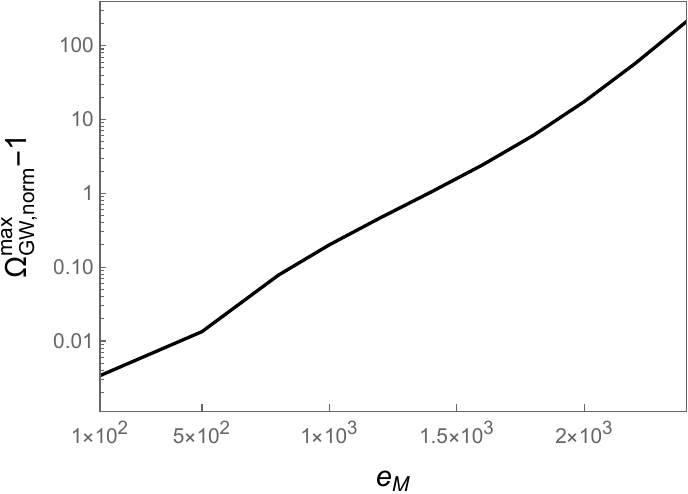}
\caption{\label{fig:GWmax2}Enhancement of the normalized 
$\Omega_{\text{GW,norm}}^{\text{max}}-1$, as a function of $\textbf{e}_M$.
The result shows that the SIGW peak amplitude is highly sensitive to the
charge-dependent mixing parameter $\textbf{e}_M$. Here $h_{\text{max}}=8$ and $\Delta=0.4$}
\end{figure}
\begin{figure}[t]
\centering
\includegraphics[scale=0.35]{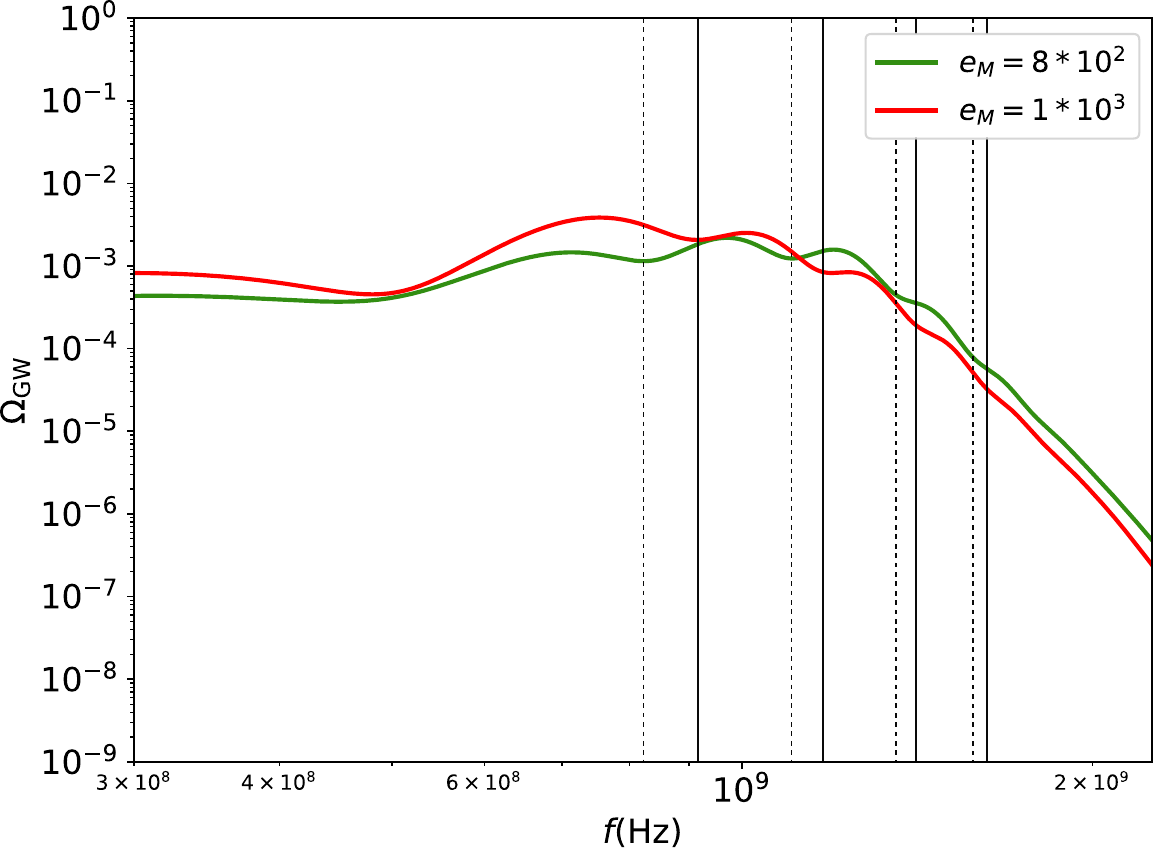}
\caption{\label{fig:GW_eM4}SIGW spectrum around the principal peak for
$\textbf{e}_M=8\times10^2$ and $\text{e}_M=1\times10^3$. The dashed and solid vertical
lines manifest the modulation pattern for $\textbf{e}_M=8\times10^2$ and
$\textbf{e}_M=1\times10^3$, respectively. Increasing $\textbf{e}_M$ shifts the modulation pattern and can move the global maximum from the second local peak to the
first local peak. Here $h_{\text{max}}=14$ and $\Delta=0.5$.}
\end{figure}
We next study the effect of coupling $\textbf{e}_M$ on the energy density of SIGWs. As we have seen in the last section, $\textbf{e}_M$ can change the frequencies of the oscillation features and the amplitudes of the peaks. This is reflected in the SIGWs, where both the peak energy density and the modulations are altered. A larger $\textbf{e}_M$ increases the peak amplitude and simultaneously widens the modulation width. The change of amplitude can be found in Fig. \ref{fig:GW_eM}. To quantify the sensitivity of the peak amplitude to the charge-dependent mixing parameter $e_M$, we analyze the variation of
$\log \Omega_{\rm GW,norm}^{\max}$ under a fractional change of $e_M$ in Fig. \ref{fig:GWmax2}. This result shows that the observable remains highly sensitive
to the parameter $\textbf{e}_M$ over the whole range of interest. For example, in the main growth region, a $10\%$ increase of $\textbf{e}_M$ typically leads to a $35\%$--$47\%$ increase of $\log \Omega^{\text{max}}_{\text{GW,norm}}$.
\begin{table*}[!t]
    \caption{Main effects of model parameters on the scalar power spectrum
    and the SIGW spectrum.}
    \label{tab:parameter_effects}
    \centering

    \begingroup
\renewcommand{\arraystretch}{1.0}
\setlength{\lightrulewidth}{0.4pt}
\setlength{\heavyrulewidth}{0.9pt}

    \begin{tabular}{@{}
        c@{\hspace{0.04\textwidth}}
        c@{\hspace{0.04\textwidth}}
        c@{}}

        \toprule

        \VCenter{0.11\textwidth}{\textbf{Parameter}}
        &
        \VCenter{0.405\textwidth}{\textbf{Main effect on \(\mathcal{P}_{\mathcal R}\)}}
        &
        \VCenter{0.405\textwidth}{\textbf{Main effect on
        \(\Omega_{\mathrm{GW}}\)}} \\

        \midrule

        \VCenter{0.11\textwidth}{\(h_{\max}\)}
        &
        \VLeft{0.405\textwidth}{
        Enhances the peak amplitude and shifts the characteristic scale,
        \(k_{\rm peak}\sim k_f h_{\max}\).}
        &
        \VLeft{0.405\textwidth}{
        Increases \(\Omega_{\rm GW}^{\max}\) and shifts the peak frequency
        \(f_{\rm peak}\).} \\

        \midrule

        \VCenter{0.11\textwidth}{\(\Delta\)}
        &
        \VLeft{0.405\textwidth}{
        Changes the width and number of oscillations.}
        &
        \VLeft{0.405\textwidth}{
        Changes the number of modulations and the degree of averaging-out.} \\

        \midrule

        \VCenter{0.11\textwidth}{\(\bm{e}_{\mu}\)}
        &
        \VLeft{0.405\textwidth}{
        Enhances the large-scale scalar spectrum and reduces the relative
        height of the main peak.}
        &
        \VLeft{0.405\textwidth}{
        Suppresses the relative principal peak and may lead to a global
        peak jump.} \\

        \midrule

        \VCenter{0.11\textwidth}{\(\bm{e}_{M}\)}
        &
        \VLeft{0.405\textwidth}{
        Modifies the peak amplitude and oscillation width.}
        &
        \VLeft{0.405\textwidth}{
        Changes \(\Omega_{\rm GW}^{\max}\) and the modulation width.} \\

        \bottomrule
    \end{tabular}
    \endgroup
\end{table*}

The shift of the modulation is shown in Fig. \ref{fig:GW_eM4}, where the modulations of SIGWs of $\textbf{e}_M=1\times10^3$ and $\textbf{e}_M=8\times10^2$ are indicated by solid and dashed vertical lines, respectively. Since increasing $\textbf{e}_M$ raises the amplitude of the large-scale peak, similar to the case of varying $\textbf{e}_{\mu}$, changing 
$\textbf{e}_M$ may also cause a jump in the global peak position. As shown in the Fig. \ref{fig:GW_eM4}, when $\textbf{e}_M$ increases from $8\times10^2$ to $1\times10^3$, the global peak shifts from the second local peak to the first local peak.

In summary, varying the charge parameters changes the relative heights of pre-existing oscillatory peaks. It provides a characteristic diagnostic of the
frequency profile of the stochastic background. This behavior is observationally useful because it can help distinguish a
charge-induced longitudinal-mode effect from a neutral sharp-feature signal. If
one only measures the peak frequency $f_{\rm peak}$, the shift may be
misinterpreted as a change in $N_f$ or $h_{\max}$. However, the simultaneous
measurement of the modulation spacing, the local peak hierarchy, and the peak
amplitude can break this degeneracy.

\section{Conclusion}\label{sec5}
In this work we have studied SIGWs generated in an inflationary model with symmetry breaking. The model contains charged scalar fields coupled to an isotropic triplet of Abelian gauge fields. Owing to the isotropic configuration of the gauge fields, the background evolution remains
compatible with an FRW universe, while the gauge-field sector can still play an important role in the dynamics of perturbations. 

We are interested in the case where a transient epoch with $h\gg1$ occurs at late times, indicating a strong mixing between inflaton perturbation and the gauge-field perturbation are strongly mixing during that period. 
The transient instability can lead to a significant peak of the curvature
power spectrum and SIGWs on small scales.

Compared to the uncharged case \cite{Chen:2023bcz}, the most significant impact is characterized by two parameters: $\textbf{e}_{\mu}$ and $\textbf{e}_M$ which respectively characterize the effects of the longitudinal mode and
charge-dependent mixing terms between perturbations in the perturbation equation. We summarize in Table \ref{tab:parameter_effects} the parameters that influence the signatures of SIGWs. The charged case modifies this universal picture in a characteristic way. The parameter $\textbf{e}_{\mu}$ mainly changes the large-scale part of the scalar spectrum and can shift the global maximum of $\Omega_{\text{GW}}$ among different local peaks, while leaving the modulation frequency almost unchanged. In contrast, $\textbf{e}_M$ can affect both the peak amplitude and the modulation width. 

        
        
        

Another characteristic feature of the charged case is the possible jump of the
global peak position. Varying $h_{\max}$ shifts the characteristic scale
approximately continuously through $k_{\rm peak}\sim k_fh_{\max}$. By contrast,
varying $\textbf{e}_\mu$ or $\textbf{e}_M$ changes the relative heights of pre-existing local
peaks in the oscillatory spectrum. Therefore, the global maximum of
$\Omega_{\rm GW}$ can jump from one local peak to another while the underlying
modulation pattern remains almost unchanged. This peak-jump behavior provides
a useful diagnostic for distinguishing the charged longitudinal sector from the
neutral transient gauge-field excitation.

The ultra-high-frequency nature of the signal has important observational implications. For the parameter choices considered in this paper, the late-time excitation places the resulting SIGWs spectrum well outside the sensitivity window of planned laser-interferometric observatories. Nevertheless, MHz--GHz band has motivated a variety of proposed high-frequency GW searches \cite{Li:2007ab,Li:2009zzy,Aggarwal:2020olq,Berlin:2021txa,Blas:2026ybh}. Although the detection of a cosmological stochastic background at these frequencies remains challenging, the characteristic spectral structures found here could, if resolved, provide distinctive signatures of the symmetry-breaking inflationary scenario.

In this paper, we only study a very toy model that is complicated (multiple scalar fields and gauge fields) but computationally
simple. We should emphasize that our discussion can be expanded to
one vector-field theory \cite{Chen:2025qyv}, where a large anisotropy is inevitable. This will be more interesting due to the gauge-field-induced anisotropy left in the stochastic gravitational wave background \cite{Chen:2022qec,Kuang:2023urj,Xie:2026xhr}. We leave all of these studies for future work.

\begin{acknowledgments}
We would like to thank Qiqi Fan for helpful discussions. This work was partly supported by  National Natural Science Foundation of China (NSFC) under Grants No. 12375049 and No. 12547125, Key Program of the Natural Science Foundation of Jiangxi Province under Grant No. 20232ACB201008, and the Ganpo High-Level Innovative Talent Program.
\end{acknowledgments}

\appendix

\section{Quadratic action}\label{qaction}

In this appendix we derive the quadratic action of perturbations we used in the text.  We first perturb the spacetime metric. Due to the isotropy of the system, we can decompose the perturbations of the spacetime as
\begin{align}\label{gravitypert}
N=&1+\alpha,\ \ \ \ N_i=\partial_i\beta+\beta_i,\ \ \ \ g_{ij}=a^2(t)(\delta_{ij}+\gamma_{ij}),\nonumber\\
\gamma_{ij}=&-2\psi\delta_{ij}+2E_{,ij}+2W_{(i,j)}+h_{ij}.
\end{align}
Here $\alpha$, $\beta$, $\psi$ and $E$ are scalar. $\beta_i$ and $W_i$ are transverse vector ($\partial^i\beta_i=\partial^iW_i=0$) and $h_{ij}$ is a transverse traceless tensor ($\partial^ih_{ij}=h^i_{\ i}=0$). 

It is known that in the spatially-flat gauge, the dominated contributions to the phenomena are the matter perturbations of the part of gauge fields \cite{Watanabe:2010fh,Emami:2013bk}. So we ignore the gravitational backreaction $(\alpha, \beta)$. After transforming the perturbations into Fourier space by replacing $\partial^2\to-k^2$, we can expand the action to second order
\begin{widetext}
\begin{align}
    \mathcal{L}_{\text{charged}}^{(2)}=\ &\frac{a^3}{2}\delta\dot{\phi}^2-\frac{a^3}{2}\left(\frac{1}{a^2}k^2+V_{\phi\phi}+\frac{1}{a^2}\text{\textbf{e}}^2A^2\right)\delta\phi^2+\frac{a^3}{6}k^2\text{\textbf{e}}^2\phi^2Y^2-\frac{a}{3}k^2\text{\textbf{e}}^2\phi^2 U^2\nonumber\\
    &-\frac{a}{6}\text{\textbf{e}}^2\phi^2\left(3\delta A^2+k^4M^2-2k^2 M\delta A\right)-\frac{2a}{3}\text{\textbf{e}}^2\phi A\left(3\delta A-k^2M\right)\delta\phi,\\
    \mathcal{L}_{\text{vector}}^{(2)}=\ &\frac{3a}{2}f^2\delta\dot{A}^2-\frac{1}{a}f^2k^2\delta A^2+af^2k^2\dot{U}^2-\frac{1}{a}f^2k^4U^2+\frac{a}{2}f^2k^4Y^2\nonumber\\
    &+\frac{a}{2}f^2k^4\dot{M}^2-af^2k^4Y\dot{M}+af^2k^2 Y\delta \dot{A}-af^2k^2\dot{M}\delta \dot{A}-2aff_{\phi}\dot{A}k^2\dot{M}\delta\phi\nonumber\\
    &+2aff_{\phi}\dot{A}\left(3\delta\dot{A}+k^2Y\right)\delta\phi+\frac{3a}{2}\left(f_{\phi}^2+ff_{\phi\phi}\right)\dot{A}^2\delta\phi^2,
\end{align}
\end{widetext}
Here we only present the scalar modes, which are decoupled with vector and tensor modes in linear order due to the isotropy.

From the equation of motion of non-dynamical mode $Y$, one can immediately obtain the following solution
\begin{align}\label{neq2}
    Y=-\frac{3f^2(\delta\dot{A}-k^2\dot{M})+6ff_{\phi}\dot{A}\delta\phi}{a^2\text{\textbf{e}}^2\phi^2+3f^2k^2}.
\end{align}
In the denominator, the second term dominates over the first term during the early stages. At later stages, the two terms in the denominator become comparable. 

After  eliminating the $Y$ we can obtain the quadratic action of all physical modes. We can further simplify its expression. 
Firstly, we note that the $U$ mode decouples with any other modes in the action. This is because $U$ is the perturbation of magnetic fields, which have vanished background value due to the isotropy and homogeneous background \cite{Firouzjahi:2018wlp,Gorji:2020vnh,Chen:2022ccf,Chen:2023bcz}. Hence we can ignore it from now on. We can also define the ``longitudinal'' mode by projecting gauge fields on the direction of wave number 
\begin{equation}
    D\equiv-\frac{k_ak_i}{k^2}A^a_{\ i}=k^2M-\delta A.
\end{equation}
Then the quadratic action is calculated to
\begin{widetext}
\begin{align}
\mathcal{L}^{(2)}_{\text{tot}}=\ &\frac{a^3}{2}\delta\dot{\phi}^2-\frac{a^3}{2}\left(\frac{1}{a^2}k^2+V_{\phi\phi}+\frac{1}{a^2}f_{\phi}^2\dot{A}^2-\frac{3}{a^2}ff_{\phi\phi}\dot{A}^2+\frac{1}{a^2}\text{\textbf{e}}^2A^2-\frac{4\text{\textbf{e}}^2\phi^2f_{\phi}^2\dot{A}^2}{3f^2k^2+a^2\text{\textbf{e}}^2\phi^2}\right)\delta\phi^2\nonumber\\
    &+af^2\delta\dot{A}^2-\frac{f^2k^2}{a}\left(1+\frac{a^2\text{\textbf{e}}^2\phi^2}{3f^2k^2}\right)\delta A^2
    +\frac{a^3\text{\textbf{e}}^2\phi^2}{2(3f^2k^2+a^2\text{\textbf{e}}^2\phi^2)}f^2\dot{D}^2
    -\frac{a\text{\textbf{e}}^2\phi^2}{6}D^2
    \nonumber\\
    &-\frac{2a^3\text{\textbf{e}}^2\phi^2}{3f^2k^2+a^2\text{\textbf{e}}^2\phi^2}ff_{\phi}\dot{A}\delta\phi\dot{D}+\frac{2a}{3}\text{\textbf{e}}^2\phi A\delta\phi D+4aff_{\phi}\dot{A}\delta\phi\delta\dot{A}-\frac{4a}{3}\text{\textbf{e}}^2\phi A\delta\phi\delta A.
\end{align}
\end{widetext}
Notably, during the early stages, the Lagrangian of the longitudinal mode is significantly smaller than that of the diagonal part of the gauge field perturbations, i.e., 
\begin{align}
    \frac{\mathcal{L}_{D^2}}{\mathcal{L}_{\delta A^2}}\sim \frac{a^2\text{\textbf{e}}^2\phi^2}{f^2k^2}\ll 1.
\end{align}
Therefore, it can be neglected at this stage. However, in later stages, it can no longer be ignored.

We can also ignore all small terms due to (\ref{neq}) and (\ref{neq2}) and all small slow-roll parameters. Then after using the background EoM in slow-roll limit \cite{Chen:2023bcz}\footnote{The minus on the r.h.s, which is different from \cite{Chen:2023bcz}, comes from the positive rolling $\dot{\phi}>0$.}
\begin{equation}\label{appA7}
    \frac{f_{\phi}}{f}M_{\text{pl}}\sqrt{2\epsilon_{\phi}}\simeq-2,\ \ \ \ \ \ \ \frac{f_{\phi\phi}}{f}\simeq\frac{f_{\phi}^2}{f^2}
\end{equation}
%
Finally, the quadratic action can be reduced to 
\begin{widetext}
\begin{align}
\label{QC}
S^{(2)}_{\text{tot}}=&\int \mathrm{d}t\mathrm{d}^3x\frac{a^3}{2}\Bigg\{\delta\dot{\phi}^2-\bigg[-\frac{\partial^2}{a^2}-8\left(1+\frac{2\lambda_{\textbf{e}}}{1+\lambda_{\textbf{e}}}\right)h^2H^2\bigg]\delta\phi^2+\delta\dot{Q}^2+\left(1+\lambda_{\textbf{e}}\right)\frac{\partial^2}{a^2}\delta Q^2\nonumber\\
&+ \delta\dot{D}^2+\bigg[\left(1+\lambda_{\textbf{e}}\right)\frac{\partial^2}{a^2}+\frac{1}{2}\frac{\ddot{\lambda}_{\textbf{e}}}{(1+\lambda_{\textbf{e}})\lambda_{\textbf{e}}}-\frac{1}{4}\frac{\dot{\lambda}_{\textbf{e}}^2}{(1+\lambda_{\textbf{e}})^2\lambda_{\textbf{e}}^2}-\frac{\dot{\lambda}_{\textbf{e}}^2}{(1+\lambda_{\textbf{e}})^2\lambda_{\textbf{e}}}-\frac{3}{2}\frac{\dot{\lambda}_{\textbf{e}}}{(1+\lambda_{\textbf{e}})\lambda_{\textbf{e}}}H\bigg] \delta D^2\nonumber\\
&+\sqrt{\frac{\lambda_{\textbf{e}}}{1+\lambda_{\textbf{e}}}}8hH\delta\phi\bigg[\delta\dot{D}+3H\delta D-\frac{1}{2}\frac{\dot{\lambda}_{\textbf{e}}}{(1+\lambda_{\textbf{e}})\lambda_{\textbf{e}}}\delta D\bigg]+\frac{4\text{\textbf{e}}^2\phi A}{3af}\sqrt{\frac{1+\lambda_{\textbf{e}}}{\lambda_{\textbf{e}}}}\delta\phi\delta D\nonumber\\
    &
+8\sqrt{2}hH\delta\phi\delta \dot{Q}
    +\left(24\sqrt{2}hH^2-\frac{4\sqrt{2}}{3af}\text{\textbf{e}}^2\phi A\right)\delta\phi\delta Q\Bigg\}.
\end{align}
\end{widetext}

\section{$\mathrm{R}$ and $\tilde{\Omega}^2$}\label{R&Omega}
The detail of the matrix $\tilde{\Omega}^2$ is shown in this appendix.
\begin{equation}
    K=\begin{pmatrix}
    0, & K_{12} & K_{13}\\
    -K_{12} &0& 0\\
    -K_{13}&0&0
    \end{pmatrix}.
\end{equation}

When $\lambda_{\textbf{e}}\ll1$, one finds $K_{13}/K_{12}\ll1$,  we can use   $\mathrm{R}'=\mathrm{R}K$, then
\begin{align}
                \mathrm{R}(\tau){}&=\exp\left(\int ^{\tau} K \mathrm d \bar\tau~\right)=\mathrm{exp}\left[\theta(\tau)\left(\frac{K}{|\boldsymbol{w}|}\right)~\right]\nonumber\\
                &=I+\left(\frac{K}{|\boldsymbol{w}|}\right)\sin\theta +\left(\frac{K}{|\boldsymbol{w}|}\right)^2(1-\cos\theta).
\end{align}
where $\theta(\tau)=\int^\tau |\boldsymbol{w}(\bar{\tau})| \mathrm{d}\bar\tau $. With $|\boldsymbol{w}|\simeq|K_{12}|$, we obtain Rodrigues’ rotation matrix 
 \begin{equation}
    \mathrm{R}(\tau)\simeq\begin{pmatrix}
\cos \theta & \frac{K_{12}}{|K_{12}|}  \sin \theta & 0 \\
-\frac{K_{12}}{|K_{12}|} \sin \theta & \cos \theta  & 0  \\
0&0  & 1 
    \end{pmatrix}.
\end{equation}
which satisfies $\mathrm{R}^{\top}\mathrm{R}=\mathbbm{1}$ and $\mathrm{R}'=\mathrm{R}K$ with
\begin{equation}
    \theta(\tau)\simeq\int_{\tau_{\text {end }}}^\tau |K_{12}| ~\mathrm{d} \bar{\tau}.
\end{equation}

With $\lambda_{\textbf{e}}\ll1$, the $\tilde{\Omega}^2$ becomes
 \begin{align}
   \tilde{\Omega}^2=\mathrm{R}\left(\Omega^2+K^{\top}K\right)\mathrm{R}^{\top}=\left(\begin{array}{ccc}
\tilde{\Omega}^2_{11} & \tilde{\Omega}^2_{12}& \tilde{\Omega}^2_{13}\\
\tilde{\Omega}^2_{12}& \tilde{\Omega}^2_{22} &\tilde{\Omega}^2_{23}  \\
\tilde{\Omega}^2_{13}& \tilde{\Omega}^2_{23} & \tilde{\Omega}^2_{33} 
\end{array}\right).
\end{align}
where
\begin{align}
\tilde{\Omega}^2_{11}\simeq{}& k^2-\frac{2}{\tau^2}+\frac{4h^2}{\tau^2}(1-\cos2\theta)\nonumber\\
&+6\sqrt{2}\left(\frac{h}{\tau^2}-\frac{a\mathbf{e}^2A\phi}{9f}\right)\sin2\theta,\\
\tilde{\Omega}^2_{22}\simeq{}& k^2-\frac{2}{\tau^2}
+\frac{4h^2}{\tau^2}(1+\cos2\theta)\nonumber\\
&-6\sqrt{2}\left(\frac{h}{\tau^2}-\frac{a\mathbf{e}^2A\phi}{9f}\right)\sin2\theta,\\
\tilde{\Omega}^2_{33}\simeq{}& k^2-\frac{2}{\tau^2}+\frac{\lambda_{\textbf{e}}(27+4h^2)}{\tau^2}.
\end{align}
and
\begin{align}
    \tilde{\Omega}^2_{12}={}&-6\sqrt{2}\left(\frac{h}{\tau^2} -\frac{a\mathbf{e}^2A\phi}{9f}\right)\cos2\theta\nonumber\\ 
    &-\frac{4h^2}{\tau^2}\sin2\theta-\frac{\lambda_{\textbf{e}} k^2}{2}\sin2\theta,
    \\
    \tilde{\Omega}^2_{13}
    \simeq{}& \frac{\sqrt{\lambda_{\textbf{e}}}}{\tau^2}\left(12h\cos\theta-4\sqrt{2}h^2\sin\theta\right)\nonumber\\
    &-\frac{2\mathbf{e} Ak}{\sqrt{3}}\cos\theta,
    \\
    \tilde{\Omega}^2_{23}
    \simeq{}&\frac{\sqrt{\lambda_{\textbf{e}}}}{\tau^2}\left(12h\sin\theta +4\sqrt{2}h^2\cos\theta\right)\nonumber\\
    &- \frac{2\mathbf{e}A k}{\sqrt{3}}\sin\theta.
\end{align}

\section{The power spectra of SIGWs}\label{app2}
 We consider the following  perturbed metric in Newtonian gauge 
\begin{equation}
\label{metric}
\mathrm{d} s^{2}
=-a^{2}(1+2 \Phi) \mathrm{d} \tau^{2}+a^{2}\left[(1-2  \Psi) \delta_{i j}+\frac12 h_{i j}\right] \mathrm{d} x^{i} \mathrm{~d} x^{j},
\end{equation} 
 where $\Phi$ and  $\Psi$ are the first-order scalar perturbations,  and  $h_{i j}$ are  the first-order and second-order tensor perturbations, respectively. Here, we neglect the anisotropic stress and it follows that $\Phi=\Psi$. \\
 
 The solution of first-order scalar perturbations during RD era in Fourier space is given by  \cite{Kohri:2018awv}
\begin{align}
\label{Phi}
    \Psi(\tau, \bm k)=& \Phi(\tau, \bm k)= \frac23\zeta_{\bm k}T_{\Phi}(x),\\
    \label{TPhi}
    \quad T_{\Phi}(x)=&\frac{9}{x^2}\left(\frac{ \sin (x/\sqrt{3})}{x/\sqrt{3}}-\cos (x/\sqrt{3})\right), 
\end{align}
where $T_{\Phi}(x)$  is the transfer function with  $x=k\tau$.

For the SIGWs, the EoM in Fourier space can be written as
\begin{equation}
\lb{h}
h_{\bm k}''+2\mathcal{H}h_{\bm k}+k^2h_{\bm k}=4  \mathcal S_{\bm k},
\end{equation}
where $\mathcal{H}\equiv aH$ and the source term is
\begin{align}
\mathcal S_{\bm k}=& \int \frac{\mathrm{d}^3p}{(2\pi)^{3/2}}\mathbf e_{ij}  p^i p^j \Big[2\Phi_{\boldsymbol{p}}\Phi_{\boldsymbol{k}-\boldsymbol{p}}\nonumber\\
&+\left(\mathcal{H}^{-1}\Phi'_{\boldsymbol{p}}+\Phi_{\boldsymbol{p}}\right) \Big(\mathcal{H}^{-1}\Phi'_{\boldsymbol{k}-\boldsymbol{p}}+\Phi_{\boldsymbol{k}-\boldsymbol{p}}\Big) 
\Big].
\end{align}
The solution of eq. \eqref{h} is 
\begin{equation}
\label{Solution}
a(\tau)h_{\bm k}\left(\tau\right)=4\int^{\tau}_{0}\mathrm{d}\bar{\tau}~G_{\bm k}\left(\tau,\bar{\tau}\right)a(\bar\tau)
\mathcal S_{\bm k}\left(\bar{\tau}\right),
\end{equation}
where   $ G_{\bm{k}}$ is the Green's function
which satisfies
$$
 G_{\bm{k}}''(\tau,\bar \tau)+\left(k^2-\frac{a''(\tau)}{a(\tau)}\right)G_{\bm{k}}=\Theta (\tau-\bar \tau)
$$
During RD era, we have 
\begin{equation}
\label{GREEN}
k  G_{\bm{k}}(\tau,\bar \tau)= \sin\left[k\left(\tau-\bar \tau \right)\right]\Theta(\tau-\bar \tau).
\end{equation}\\

The power spectra $\mathcal{P}_{h}(k,\tau)$ are related to the expectation values as
\begin{equation}
\label{hh}
   \left \langle h^{\lambda}_{\bm{k}}(\tau) h^{\lambda'}_{\bm{k}'}(\tau)\right\rangle =\delta^{\lambda \lambda'}\delta^3(\bm k+\bm k')\frac{2\pi^2}{k^3}\mathcal{P}_{h}(k,\tau).
\end{equation}
where  $\lambda,\lambda'=+,\times$ represents the polarization index, which we omit in the following.

\bibliographystyle{apsrev4-1}
\bibliography{ref}
\end{document}